\documentclass[preprint,authoryear,x11names]{imsart}

\RequirePackage[OT1]{fontenc}
\RequirePackage{amsthm,amsmath,amssymb,bm,dsfont}
\RequirePackage[numbers]{natbib}
\RequirePackage[colorlinks,citecolor=blue,urlcolor=blue]{hyperref}

\usepackage[left=3.75cm,right=3.75cm]{geometry}

\startlocaldefs
\numberwithin{equation}{section}
\theoremstyle{plain}

\endlocaldefs

\usepackage{amsmath,amssymb,bm,dsfont,fontawesome}
\usepackage{mathtools}
\mathtoolsset{showonlyrefs}

\usepackage[ruled,linesnumbered]{algorithm2e}
\let\oldnl\nl
\newcommand{\nonl}{\renewcommand{\nl}{\let\nl\oldnl}}

\renewcommand{\vec}{\boldsymbol}
\newcommand{\mvec}{\mathbf}
\newcommand{\abs}[1]{\vert{#1}\vert}

\usepackage[misc]{ifsym}
\usepackage{dutchcal,xcolor,verbatim}
\newcommand{\nlsum}{\sum}
\newcommand{\nlprod}{\prod}
\newcommand{\gt}{t^\prime}
\renewcommand{\top}{\intercal}

\usepackage{booktabs}	
\usepackage{array,tabu}
\usepackage{multicol}
\usepackage{multirow}
\usepackage{siunitx}

\usepackage{subcaption}
\usepackage[figurename=Fig, font=footnotesize, labelfont=sc, textfont=it]{caption}

\usepackage{numprint}
\npthousandsep{,}\npthousandthpartsep{}\npdecimalsign{.}

\begin{document}

\begin{frontmatter}
\title{Graph link prediction in computer networks using Poisson matrix factorisation
}
\runtitle{Graph link prediction in computer networks using PMF}

\begin{aug}
\author{\fnms{Francesco} \snm{Sanna Passino}$\color{blue}{}^\dagger$},
\author{\fnms{Melissa J.\hspace{0.1em}M.} \snm{Turcotte}$\color{blue}{}^\ddagger$\thanksref{t1}}, \\
\and
\author{\fnms{Nicholas A.} \snm{Heard}$\color{blue}{}^\dagger$} \\

\runauthor{Sanna Passino, Turcotte and Heard}

\address{
$\color{blue}\dagger$ -- Department of Mathematics, Imperial College London \\
$\color{blue}\ddagger$ -- Advanced Research in Cyber-Systems, Los Alamos National Laboratory
}
\thankstext{t1}{The author is currently at Microsoft 365 Defender, Microsoft Corporation. This work was completed while the author was at the Los Alamos National Laboratory.}

\end{aug}

\begin{abstract}
Graph link prediction is an important task in cyber-security: relationships between entities within a computer network, such as users interacting with computers, or system libraries and the corresponding processes that use them, can provide key insights into adversary behaviour.
  Poisson matrix factorisation (PMF) is a popular model for link prediction in large networks, particularly useful for its scalability.
  In this article, PMF is extended to include scenarios that are commonly encountered in cyber-security applications. 
  Specifically, an extension is proposed to explicitly handle binary adjacency matrices and include known categorical covariates associated with the graph nodes. 
 A seasonal PMF model is also presented to handle seasonal networks. To allow the methods to scale to large graphs, variational methods are discussed for performing fast inference.
  The results show an improved performance over
  the standard PMF model and other statistical network models.
\end{abstract}



\end{frontmatter}

\section{Introduction} \label{intro}
In recent years, there has been a significant increase in investment from both government and industry in improving cyber-security using statistical and machine learning techniques on a wide range of data
collected from computer networks \citep{HeardAdams18, Jeske18}.
One significant research challenge associated with these networks is {\it link prediction}, defined as the problem of
predicting the presence of an edge between two nodes in a network graph, based on observed edges and attributes of the nodes \citep{liben}. Adversaries attacking a computer network often affect relationships (links) between nodes within these networks, such as users authenticating to computers, or clients connecting to servers. \textit{New links} (previously unobserved relationships) are of particular interest, as many attack behaviours such as lateral movement \citep{neil}, phishing, and data retrieval, can create new edges between network entities \citep{Metelli19}. In practical cyber applications, it is necessary to use relatively simple and scalable statistical methods, given the size and inherently dynamic nature of these networks.

Away from cyber applications, the link
prediction problem has been an active field of research \citep[see, for example,][]{dunlavy,lu,menon}, being similar, especially in its \textit{static} formulation, to recommender systems \citep{adomavicius}. 
Static link prediction \citep[for example,][]{Clauset08} aims at filling in missing entries in a single incomplete graph adjacency matrix, as opposed to temporal link prediction \citep[for example,][]{dunlavy}, which aims at predicting future snapshots of the graph given one or more fully observed snapshots. Static link prediction problems have been successfully tackled using probabilistic matrix factorisation methods, especially classical Gaussian matrix factorisation \citep{prob_mat_fact}, and are currently widely used in the technology industry \citep[see, for example,][]{agarwal_aoas,Khanna13,Paquet13}. 
For dyadic count data, Poisson matrix factorisation (PMF) \citep{canny,dunson,cemgil,gopalan} emerged as a suitable model in the static link prediction framework. 
This work mainly focuses on temporal link prediction, showing that PMF is also useful for link prediction in this context, and it seems to be particularly well-suited for cyber-security applications.

The methodological contribution of this article is to present extensions of the PMF model, suitably adapted to scenarios which are commonly encountered in cyber-security computer network applications.
Traditionally, Poisson matrix factorisation methods are used on partially observed adjacency matrices of natural numbers representing, for example, ratings of movies provided by different users. In computer networks, the matrix is fully observed, and the counts associated with network edges are complicated by repeated observations, polling at regular intervals, and the intrinsic burstiness of the events \citep{Heard14}. Hence, each edge is usually represented by a binary indicator expressing whether at least one connection between the corresponding nodes was observed.
Consequently, the standard PMF model, where counts associated with links are assumed to follow a Poisson distribution with unbounded support, cannot be applied directly. Instead, indicator functions are applied, leading to an extension for PMF on binary adjacency matrices.
Next, a framework for including categorical covariates within the PMF model is introduced, which also allows for modelling of new nodes appearing within a network.
Finally, extensions of the PMF model to incorporate seasonal dynamics are presented.

The rest of the article is organised as follows: Section~\ref{sec:data} presents the computer network data which are to be analysed. Section~\ref{background} formally introduces Poisson matrix factorisation for network link prediction, and Section~\ref{pmf_covs} discusses the proposed PMF model for binary matrices and labelled nodes.
A seasonal extension is described in Section~\ref{seasonal_pmf}. Finally, results of the analysis are presented in Section~\ref{results_section}.

\section{LANL computer network data}\label{sec:data}
The methodologies in this article have been developed to provide insight into authentication data extracted from the publicly released ``Unified Host and Network Dataset'' from Los Alamos National Laboratory (LANL) \citep{Turcotte18}.

The data contain authentication logs collected over a 90-day period from computers in the Los Alamos National Laboratory enterprise network running a Microsoft Windows operating system. An example record is: 

\begingroup\small
\begin{verbatim}
{"UserName": "User865586", "EventID": 4624, "LogHost": "Comp256596", 
 "LogonID": "0x5aa8bd4", "DomainName": "Domain001", 
 "LogonTypeDescription": "Network", "Source": "Comp782342", 
 "AuthenticationPackage": "Kerberos", "Time": 87264, "LogonType": 3}.
\end{verbatim}
\endgroup

From each authentication record, the following fields are extracted for analysis: the user credential that initiated the event ({\tt UserName}), the computer where the authentication originated ({\tt Source}), and the computer the credential was authenticating to (most often {\tt LogHost}). 
Two bipartite graphs are then generated: first, the network users and the computers from which they authenticate, denoted \textit{User -- Source}; second, the same users and the computers or servers they are connecting to, denoted \textit{User -- Destination}. The two graphs are first analysed separately, before a joint model is explored.

As generic notation, let $\mathbb G=(U,V,E)$ represent one of these bipartite graphs, 
where $U=\{u_1,u_2,\ldots\}$ is the set of {\it users} and $V=\{v_1,v_2,\ldots\}$ a set of {\it computers} (sometimes referred to as {\it hosts}). The set $E\subseteq U\times V$ represents the observed edges, such that $(u,v)\in E$ if user $u\in U$ connected to host $v\in V$ in a given time interval.
A finite set of edges $E$ can be represented as a rectangular $\abs{U}\times\abs{V}$ binary adjacency matrix $\mvec A$, where $A_{ij} =\mathds 1_E\{(u_i,v_j)\}$.

Importantly, some additional node-level categorical covariates were also available for this analysis. 
Six covariates were obtained for the users, corresponding to position within the hierarchy of the organisation as well as location and job category. For the computer hosts, three covariates were available, relating to the machine type, subnet and location within the organisation. To preserve privacy these covariates are anonymised in this study.

In total, there are $K=\numprint{1064}$ factor levels available for the user credentials and $H=735$ factor levels for the computers. One objective of this article is to present methodology for incorporating such covariates within Poisson matrix factorisation.
Within cyber-security, the availability of the LANL network with covariates is particularly significant, since the lack of appropriate datasets has been identified as one of the main limitations to widespread applications of data science methods to cyber-security problems \citep{Kumar17,Anderson18,Amit19}. 

\begin{figure}[!t]
\centering
\includegraphics[width=\textwidth]{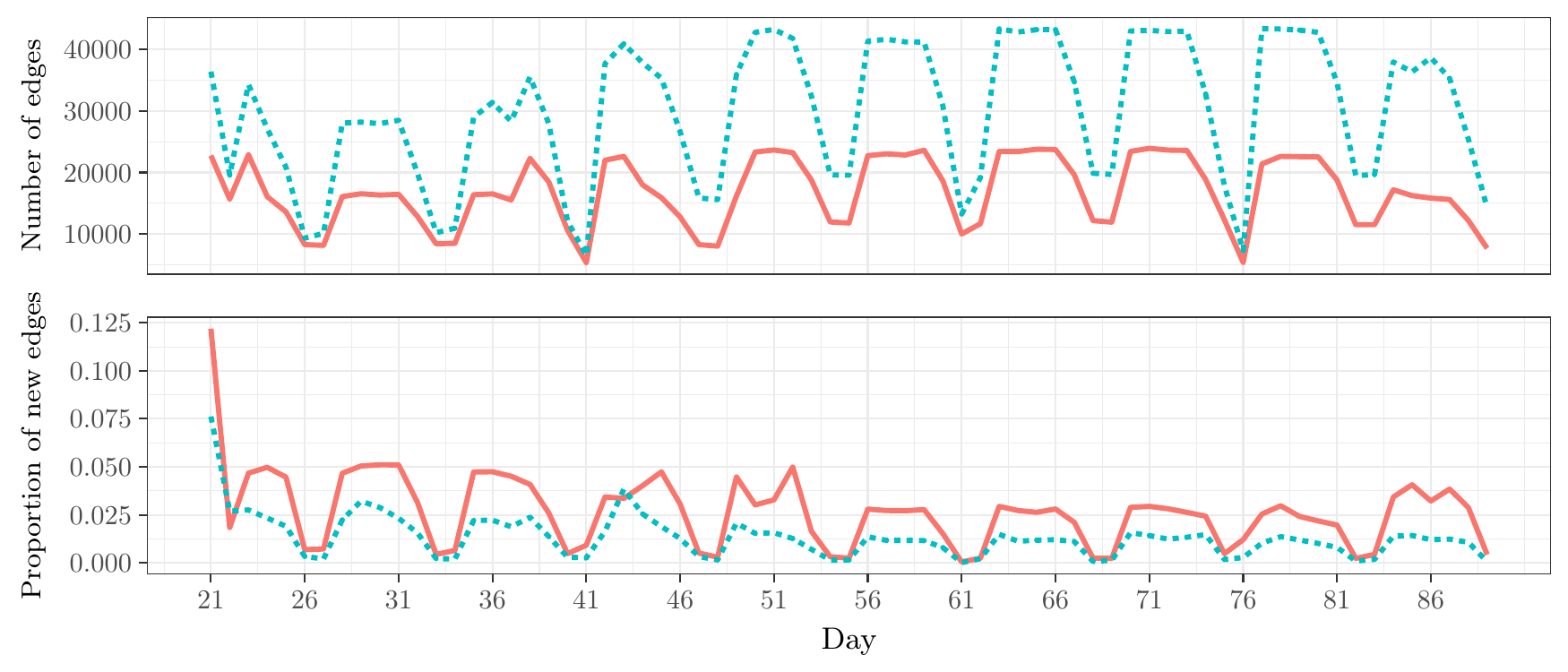}
\definecolor{corallo}{RGB}{0,191,196}
\definecolor{corallo2}{RGB}{248,118,109}
\caption{Number of links per day (top), and proportion of those that are new (bottom), after $20$ days of observation of the LANL computer network. {\bf\color{corallo2} Solid red} curve: \textit{User -- Source}. {\bf\color{corallo} Dashed blue} curve: \textit{User -- Destination}.}
\label{new_links_plot}
\end{figure}

As mentioned in Section~\ref{intro}, for cyber-security applications it would be valuable to accurately predict and assess the significance of new links. Importantly, many new links are formed each day as part of normal operating behaviour of a computer network; to demonstrate this, Figure~\ref{new_links_plot} shows the the total number of edges formed each day and the 
proportion of those that are new for the \textit{User -- Source} and \textit{User -- Destination} graphs. Even though the relative percentage is small, this would still provide many more alerts than could be practically acted upon each day.

\section{Background on Poisson matrix factorisation} \label{background}

Let $\mvec N\in\mathbb N^{\abs{U}\times\abs{V}}$ be a matrix of non-negative integers $N_{ij}$. 
For recommender system applications, $N_{ij}$ could represent information about how a user $i$ rated an item $j$, or a count of the times they have clicked on or purchased the item.
The cyber-security application has a major difference: in recommender systems, if user $i$ never interacted with the item $j$, then $N_{ij}$ is considered as a \textit{missing observation}, implying that the that the number of observations is a possibly very small fraction of $\abs{U}\abs{V}$. On the other hand, in cyber-security, the absence of a link from $i$ to $j$ is itself an observation, namely $N_{ij}=0$, and $\abs{U}\abs{V}$ data points are observed. Such a difference has practical consequences, particularly regarding scalability in cyber-security of models borrowed from the recommender systems literature. 

The hierarchical Poisson factorisation model \citep{gopalan} specifies a distribution for 
$N_{ij}$ using a Poisson link function with rate given by the inner product between user-specific latent features $\bm\alpha_i\in\mathbb R^R_+$
and host-specific latent features $\bm\beta_j\in\mathbb R^R_+$, for a positive integer $R\geq 1$:
\begin{equation}
N_{ij}
\sim\mathrm{Pois}(\bm\alpha_i^\top\bm\beta_j)=\mathrm{Pois}\left(\nlsum_{r=1}^R \alpha_{ir}\beta_{jr}\right). \label{pmf_equation}
\end{equation}
If two latent features are \textit{close} in the latent space, the corresponding nodes are expected to exhibit similar connectivity patterns. 
The specification of the model is completed 
with gamma hierarchical priors: 
\begin{align}
&\alpha_{ir} \sim\Gamma(a^{(\alpha)},\zeta_i^{(\alpha)}), \ i=1,\dots,\abs{U},\ r=1,\dots,R, \\
&\beta_{jr}\sim\Gamma(a^{(\beta)},\zeta_j^{(\beta)}), \ j=1,\dots,\abs{V},\ r=1,\dots,R, \\
& \zeta_i^{(\alpha)}\sim\Gamma(b^{(\alpha)},c^{(\alpha)}),\ \zeta_j^{(\beta)}\sim\Gamma(b^{(\beta)},c^{(\beta)}). 
\label{pmf_standard}
\end{align}
Each of the gamma distribution parameters $a^{(\alpha)}, b^{(\alpha)}, c^{(\alpha)}, a^{(\beta)}, b^{(\beta)}, c^{(\beta)}$ are positive real numbers which must be specified.

In the cyber-security context there are no missing observations in the matrix $\mvec N$. Consequently, an advantage of PMF over other models for link prediction \citep[for example,][]{prob_mat_fact} is that the likelihood function only depends on the observed links, meaning evaluating the likelihood is $\mathcal O(\mathrm{nnz}(\mvec N))$, where $\mathrm{nnz}(\cdot)$ is the number of non-zero elements in the matrix,
compared to $\mathcal O(\abs{U}\abs{V})$ for most statistical network models. 
In cyber-security, networks tend to be very large in the number of nodes, but extremely sparse:
$\mathrm{nnz}(\mvec N)\ll \abs{U}\abs{V}$. Hence, PMF appears to be a particularly appealing modelling framework for this application.

The PMF model has been used as a building block for multiple extensions. 
For example, \cite{chaney} developed \textit{social Poisson factorisation} to include latent \textit{social influences} in personalised recommendations. 
\cite{gopalan_cpf} developed \textit{collaborative topic Poisson factorisation}, which adds a document topic offset to the standard PMF model to provide content-based recommendations and thereby tackle the challenge of recommending new items, referred to in the literature as cold starts. These ideas of combining collaborative filtering and content-based filtering are further developed in \cite{Zhang15}, \cite{singh} and \cite{da_silva}, where social influences are added as constraints in the latter. 

In general, most PMF-based methods presented in this section model \textit{binary} adjacency matrices 
using the Poisson link function for convenience. This approach is computationally advantageous, but implies an incorrect model for the range: the entries $A_{ij}$ of the adjacency matrix are binary, whereas the Poisson distribution has support over the natural numbers.

It must be noted that many other models besides PMF have been proposed in the literature to tackle the link prediction task and for graph inference. Comprehensive surveys of the most popular statistical network models are given in \cite{Goldenberg10} and \cite{Fienberg12}. The problem of link prediction is also studied in other disciplines, for example physics \citep{lu}, or computer science, where graph neural networks \citep{Zhang18,Wu20} have recently gained popularity. The model presented in this article can be classified as a latent feature model \citep[LFM,][]{hoff}. 
For a bipartite graph $\mathbb G=(U,V,E)$ with adjacency matrix $\mvec A$, the LFM assumes that the nodes have $R$-dimensional latent representations $\vec u_i\in\mathbb R^R,\ i\in U$ and $\vec v_j\in\mathbb R^R,\ j\in V$. The entries of the adjacency matrix are then obtained independently as $\mathbb P(A_{ij}=1)=\kappa(\vec u_i,\vec v_j)$, where $\kappa:\mathbb R^R\times\mathbb R^R\to[0,1]$ is a kernel function. A popular special case of LFM is the random dot product graph \citep[RDPG,][]{Athreya18}, where $\kappa(\cdot)$ is chosen to be the inner product between the latent positions. The extended PMF 
model proposed in this article is also a special case of LFM, assuming a particular form of kernel function with nodal covariates. 

Dynamical extensions to PMF have also been studied. 
\cite{charlin} use Gaussian random walk updates on the latent features to dynamically correct the rates of the Poisson distributions. 
\cite{schein,schein2} propose a temporal version of PMF using the two main tensor factorisation algorithms: canonical polyadic and Tucker decompositions. 
\cite{Hosseini18} combine the PMF model with the Poisson process to produce dynamic recommendations. 
Dynamic network models have also been widely studied outside the domain of matrix factorisation techniques \citep[for a survey, see][]{Kim17}. For example, \cite{Sewell15} extend LFMs to a temporal setting.
The dynamic models described above could handle generic network dynamics, but seasonality, a special case of temporal structure, has not been explicitly accounted for in the PMF literature.
This article further aims to fill this gap and propose a viable seasonal PMF model.

\section{PMF with labelled nodes and binary adjacency matrices} \label{pmf_covs}

Suppose that there are $K$ covariates associated with each user and $H$ covariates for each host.
Let the value of the covariate $k$ for user $i$ be denoted as $x_{ik}$. Similarly, let the value of the covariate $h$ for host $j$ be $y_{jh}$. In cyber-security applications, most available information types are categorical, indicating memberships of known groupings or clusters of nodes. For the remaining of
this article, the covariates will be assumed to be binary indicators representing categorical variables. This type of encoding of categorical variables is commonly known in the statistical literature as \textit{dummy variable encoding}, whereas the term \textit{one-hot encoding} is used in the machine learning community. Several approaches for including nodal covariates in recommender systems using non-probabilistic matrix factorisation methods such as the Singular Value Decomposition (SVD) have been discussed in the literature \citep[for some examples, see][]{Nguyen13,Fithian18,Dai19}. Regression methods for network data with covariates are also studied in \cite{hoff_bilinear}. 
In Section~\ref{intro} and~\ref{background}, it has been remarked that, for binary adjacency matrices, the standard PMF model for count data cannot be applied directly, since the observations are binary, whereas the Poisson distribution has support on the natural numbers. 
To model binary links, it is assumed here that the count $N_{ij}$ is a latent random variable, and the binary indicator $A_{ij}=\mathds 1_{\mathbb N_+}(N_{ij})$ is a censored Poisson draw with a corresponding Bernoulli distribution. This type of link has been referred to in the literature as the Bernoulli-Poisson (BerPo) link \citep{acharya,zhou}. 
The full extended model is
\begin{align}
 A_{ij}\vert N_{ij} &=\  \mathds 1_{\mathbb N_+}(N_{ij}),\\
  \ N_{ij}\vert\bm\alpha_i,\bm\beta_j,\bm\Phi &\sim  \mathrm{Pois}\left(\bm\alpha_i^\top\bm\beta_j+\vec 1_K^\top(\bm\Phi\odot\vec x_i\vec y_j^\top)\vec 1_H\right) \\
&= \mathrm{Pois}\left(\nlsum_{r=1}^R\alpha_{ir}\beta_{jr}+\nlsum_{k=1}^K\nlsum_{h=1}^H\phi_{kh}x_{ik}y_{jh}\right), \label{nij}
\end{align}
where $\vec 1_n$ is a vector of $n$ ones, $\odot$ is the Hadamard element-wise product, and $\vec x_i$ and $\vec y_j$ are $K$ and $H$-dimensional binary vectors of covariates. The $R$-dimensional latent features $\bm\alpha_i$ and $\bm\beta_j$ appear in the traditional PMF model given in \eqref{pmf_equation}, and $\bm\Phi=\{\phi_{kh}\}\in\mathbb R^{K\times H}_+$ is a matrix of interaction terms for each combination of the covariates. Under model \eqref{nij},
\begin{equation}
{\mathbb P}(A_{ij}=1)=1- \exp\left(-\nlsum_{r=1}^R\alpha_{ir}\beta_{jr}-\nlsum_{k=1}^K\nlsum_{h=1}^H\phi_{kh}x_{ik}y_{jh}\right).\label{eq:paij}
\end{equation}

To provide intuition for these extra terms, assume for the cyber-security application that a binary covariate
for job title {\tt manager} is provided for the users, and that a binary covariate for the location {\tt research lab} for the hosts. If user $i$ is a manager and host
$j$ is located in a research lab, then $\phi_{kh}$ expresses a correction to the rate $\bm\alpha_i^\top\bm\beta_j$ for a manager connecting to a machine in a research lab. The covariate term is inspired by the bilinear mixed-effects models for network data in \cite{hoff_bilinear}. 

The same hierarchical priors \eqref{pmf_standard} are used for $\bm\alpha_i$ and $\bm\beta_j$ and the following prior distribution completes the specification of the model:
\begin{align}
\phi_{kh}\vert\zeta^{(\phi)} &\sim\Gamma(a^{(\phi)},\zeta^{(\phi)}),\ k=1,\dots,K,\ h=1,\dots,H,\\ \zeta^{(\phi)}&\sim\Gamma(b^{(\phi)},c^{(\phi)}). \notag
\end{align}

Note that this model provides a natural way for handling what the literature commonly refers to as cold starts, where new users or hosts appear in the network. Provided that covariate-level information is known about new entities, then the estimates for $\bm \Phi$ can be used to make predictions about links where $\bm \alpha_i$ and $\bm \beta_j$ for new user $i$ or new host $j$ could be initialised from the prior or some other global statistic based on other users and hosts.

Another significant advantage of the model proposed in \eqref{nij} is that the likelihood is $\mathcal O(\mathrm{nnz}(\mvec A))$, analogously to the standard PMF model in \eqref{pmf_standard}, implying that the model scales well to large sparse networks. From \eqref{eq:paij}:
\begin{equation}
\log L(\mvec A) = \sum_{i,j:A_{ij}>0} \log\left(e^{\psi_{ij}}-1\right) - \left(\nlsum_i\bm\alpha_i\right)^\top\left(\nlsum_j\bm\beta_j\right) - \nlsum_{k,h}^{K,H} \phi_{kh} \tilde x_k\tilde y_h, \notag
\end{equation}
where $\psi_{ij}=\bm\alpha_i^\top\bm\beta_j+\vec 1_K^\top(\bm\Phi\odot\vec x_i\vec y_j^\top)\vec 1_H$, $\tilde x_k=\sum_{i=1}^{\abs{U}} x_{ik}$, and $\tilde y_h=\sum_{j=1}^{\abs{V}} y_{jh}$.

\subsection{Bayesian inference} \label{bayes_inference}

Given an observed matrix $\mvec A$, inferential interest is on the marginal posterior distributions of the parameters $\bm\alpha_i$ and $\bm\beta_j$
for all the users and hosts, and the 
parameters $\bm\Phi$ for the covariates, since these govern the predictive distribution for the edges observed in the future.

A common approach for performing inference is adopted, where additional latent variables are introduced. Given the (assumed) unobserved count $N_{ij}$, a further set of latent counts $Z_{ijl},\ l=1,\dots,R+KH$, are used to represent the contribution of each component $l$ to the total latent count, such that $N_{ij}=\sum_{l} Z_{ijl}$.
For $l\leq R$, $Z_{ijl}\sim\mathrm{Pois}(\alpha_{il}\beta_{jl})$. Otherwise, $l$ refers to a $(k,h)$ covariate pair, and
$Z_{ijl}\sim\mathrm{Pois}(\phi_{kh})$. This construction ensures that $N_{ij}$ has precisely
the Poisson distribution 
specified in \eqref{nij}. 

Inference using Gibbs sampling is straightforward, as the full conditionals all have closed form expressions, but sampling-based methods do not scale
well with network size. Instead, a variational inference procedure is proposed. Variational inference schemes have already been commonly and successfully used in the literature for network models 
\citep{Nakajima10,Seeger12,Salter13,Lobato14}, despite the issue of introducing bias and potentially reducing estimation accuracy, in particular on the posterior variability \citep[see, for example,][]{Huggins19}. 

Gibbs sampling also presents an additional difficulty in PMF models: the inner product $\bm\alpha_i^\top\bm\beta_j$ is invariant to 
permutations of the latent features. In particular, $(\mvec Q\bm\alpha_i)^\top(\mvec Q\bm\beta_j)=\bm\alpha_i^\top\bm\beta_j$ for any 
permutation matrix $\mvec Q\in\mathbb R^{R\times R}$, which makes the 
posterior invariant under such transformation. This implies that the posterior is highly multimodal, a well-known burden for MCMC-based inference in Bayesian factor models \citep{Papastamoulis20}. 
Hence, parameter estimates obtained as averages from MCMC samples from the posterior could be meaningless, since the algorithm could have switched among different modes. 
On the other hand, variational inference is well understood to be a \textit{``mode-seeking"} algorithm, 
a desirable property for this problem. 
A practical comparison of the estimates of the inner products for prediction purposes will be briefly illustrated in Section~\ref{gibbs_results}.

\subsection{Variational inference} \label{inference}

Variational inference \citep[see, for example,][]{blei} is an optimisation based technique for approximating intractable distributions, such as the joint posterior density $p(\bm\alpha,\bm\beta,\bm\Phi,\bm\zeta,\mvec N,\mvec Z\vert\mvec A)$,
with a proxy $q(\bm\alpha,\bm\beta,\bm\Phi,\bm\zeta,\mvec N,\mvec Z)$ from a given distributional family $\mathcal{Q}$, and then finding the member $q^\star\in\mathcal{Q}$ that minimises the Kullback-Leibler (KL) divergence to the true posterior. Usually the KL-divergence cannot be explicitly computed, and therefore an equivalent objective, called the {\it evidence lower bound} (ELBO), is maximised instead:
\begin{equation}
\mathrm{ELBO}(q) = \mathbb E^q[\log p(\bm\alpha,\bm\beta,\bm\Phi,\bm\zeta,\mvec N,\mvec Z,\mvec A)]- \mathbb E^q[\log q(\bm\alpha,\bm\beta,\bm\Phi,\bm\zeta,\mvec N,\mvec Z)], \label{elbo}
\end{equation}
where the expectations are taken with respect to $q(\cdot)$.
The proxy
distribution $q(\cdot)$ is usually chosen to be of much simpler form than the posterior distribution, so that maximising the ELBO is tractable. As in \cite{gopalan} the {\it mean-field variational family} is used, where the latent variables in the posterior are considered to be independent and governed by their own distribution, so that:
\begin{gather}
q(\bm\alpha,\bm\beta,\bm\zeta,\bm\Phi,\mvec N,\mvec Z) = \nlprod_{i,r} q(\alpha_{ir}\vert\lambda_{ir}^{(\alpha)},\mu_{ir}^{(\alpha)})\times\nlprod_{j,r}q(\beta_{jr}\vert\lambda_{jr}^{(\beta)},\mu_{jr}^{(\beta)}) \\ 
\times\nlprod_{k,h}q(\phi_{kh}\vert\lambda_{kh}^{(\phi)},\mu_{kh}^{(\phi)})\times\nlprod_i q(\zeta_i^{(\alpha)}\vert\nu_i^{(\alpha)},\xi_i^{(\alpha)}) \times\nlprod_j q(\zeta_j^{(\beta)}\vert\nu_j^{(\beta)},\xi_j^{(\beta)}) \\ 
\times q(\zeta^{(\phi)}\vert\nu^{(\phi)},\xi^{(\phi)})\times\nlprod_{i,j}q(N_{ij},\mvec Z_{ij}\vert\theta_{ij},\bm\chi_{ij}). \label{mean_field}
\end{gather} 
The objective function \eqref{elbo} is optimised using {\it coordinate ascent mean field variational inference} (CAVI), whereby each density or variational factor is optimised while holding the others fixed \citep[see][for details]{bishop, blei}.
Using this algorithm the optimal form of each variational factor is: 
\begin{equation}
q^\star(v_j)\propto\exp\left\{\mathbb E^q_{-j}\left[\log p(v_{j}\vert\vec v_{-j},\mvec A)\right]\right\}, \label{cavi_update}
\end{equation}
where $v_j$ is an element of a \textit{partition} of the full set of parameters $\vec v$, and the expectation is taken with respect to the variational densities that are currently held fixed for $\vec v_{-j}$, defined as $\vec v$ excluding the parameters in the subset $v_j$. Convergence of the CAVI algorithm is determined by monitoring the change in the ELBO over subsequent iterations.

Since the prior distributions are chosen to be conjugate, the full conditionals in \eqref{cavi_update} are all available analytically. Full details are given in Appendix~\ref{supp_full}. Also, 
since all the conditionals are exponential families, each $q(v_j)$ obtained from \eqref{cavi_update} is from the same exponential family \citep{blei}.
Hence, with the exception of $q(N_{ij},\mvec Z_{ij}\vert\theta_{ij},\bm\chi_{ij})$, the proxy distributions in \eqref{mean_field} are all gamma; for example,  $q(\alpha_{ir}\vert\lambda_{ir}^{(\alpha)},\mu_{ir}^{(\alpha)})=\Gamma(\lambda_{ir}^{(\alpha)},\mu_{ir}^{(\alpha)})$.
The update equations for the 
parameters $\{\bm\lambda, \bm\mu, \bm\nu, \bm\xi, \bm\theta, \bm\chi\}$ 
of the variational approximation can be obtained using \eqref{cavi_update}, which is effectively the expected parameter of the full conditional with respect to $q$. 
The full variational inference algorithm is detailed in Algorithm~\ref{algo_vi}.  Note that each update equation only depends upon the elements of the matrix where $A_{ij} > 0$, providing computational efficiency for large sparse matrices.
Further details concerning the update equations for the Poisson and multinomial parameters $\bm\theta$ and $\bm\chi$ are also given in Appendix~\ref{supp_vi}. 

\begin{algorithm}[p]
\normalsize
\SetAlgoLined
initialise $\bm\lambda,\bm\mu$
and $\bm\xi$ from the prior, 
\\
set 
$\nu_i^{(\alpha)}=b^{(\alpha)}+Ra^{(\alpha)},\ \nu_j^{(\beta)}=b^{(\beta)}+Ra^{(\beta)},\ \nu^{(\phi)}=b^{(\phi)}+KHa^{(\phi)}$, \\
calculate $\tilde x_k=\sum_{i=1}^{\abs{U}}x_{ik},\ k=1,\dots,K$ and $\tilde y_h=\sum_{j=1}^{\abs{V}}y_{jh}$, \\
 \Repeat{convergence}{
  for each entry of $\mvec A$ such that $A_{ij}>0$, update the rate $\theta_{ij}$:
   \begin{align}
   \theta_{ij} =&\ \nlsum_{r=1}^R \exp\left\{\Psi(\lambda_{ir}^{(\alpha)}) - \log(\mu_{ir}^{(\alpha)}) + \Psi(\lambda_{jr}^{(\beta)}) - \log(\mu_{jr}^{(\beta)})\right\} \\
   			&+\nlsum_{k=1}^K\nlsum_{h=1}^H x_{ik}y_{jh}\exp\left\{\Psi(\lambda_{kh}^{(\phi)})-\log(\mu_{kh}^{(\phi)})\right\},
 \label{update_thetaij}
 \end{align} 
 where $\Psi(\cdot)$ is the digamma function, \\
  for each entry of $\mvec A$ such that $A_{ij}>0$, update $\chi_{ijl}$:  \nonl 
  \begin{align} 
  \chi_{ijl}\propto\left\{ \begin{array}{ll} \exp\left\{\Psi(\lambda_{il}^{(\alpha)})-\log(\mu_{il}^{(\alpha)})+\Psi(\lambda_{jl}^{(\beta)})-\log(\mu_{jl}^{(\beta)})\right\} & l\leq R, \\
  x_{ik}y_{jh}\exp\left\{\Psi(\lambda_{kh}^{(\phi)})-\log(\mu_{kh}^{(\phi)})\right\} & l>R, 
  \end{array}\right. \label{update_chi}\notag
  \end{align} \\ 
  \hspace*{2pt} where, for $l>R$, $l$ corresponds to a covariate pair $(k,h)$, \\ 
 
  update the user-specific first-level parameters: 
  \begin{equation}
 \lambda_{ir}^{(\alpha)} = a^{(\alpha)} +\sum_{j=1}^{\abs{V}} \frac{A_{ij}\theta_{ij}\chi_{ijr}}{1-e^{-\theta_{ij}}},\
\mu_{ir}^{(\alpha)} =  \frac{\nu_i^{(\alpha)}}{\xi_i^{(\alpha)}} + \sum_{j=1}^{\abs{V}} \frac{\lambda_{jr}^{(\beta)}}{\mu_{jr}^{(\beta)}},\
 \label{update_vi_lambda}
  \end{equation} 
  
  update the host-specific first-level parameters:
  \begin{equation}
 \lambda_{jr}^{(\beta)} = a^{(\beta)} +\sum_{i=1}^{\abs{U}} \frac{A_{ij}\theta_{ij}\chi_{ijr}}{1-e^{-\theta_{ij}}}, \
\mu_{jr}^{(\beta)} = \frac{\nu_j^{(\beta)}}{\xi_j^{(\beta)}} + \sum_{i=1}^{\abs{U}} \frac{\lambda_{ir}^{(\alpha)}}{\mu_{ir}^{(\alpha)}}, \
 \label{update_vi_mu}
  \end{equation}
   
    update the covariate-specific first-level parameters:
  \begin{equation} 
\lambda_{kh}^{(\phi)} = a^{(\phi)} + \sum_{i,j=1}^{\abs{U},\abs{V}} \frac{A_{ij}\theta_{ij}\chi_{ijl}}{1-e^{-\theta_{ij}}},\ 
\mu_{kh}^{(\phi)} = \frac{\nu^{(\phi)}}{\xi^{(\phi)}} + \tilde x_k\tilde y_h,\
\label{update_vi_covs}
  \end{equation} 
     
    update the second-level parameters:
  \begin{equation} 
\xi_i^{(\alpha)} = c^{(\alpha)} + \sum_{r=1}^R \frac{\lambda_{ir}^{(\alpha)}}{\mu_{ir}^{(\alpha)}},\
\xi_j^{(\beta)} = c^{(\beta)} + \sum_{r=1}^R \frac{\lambda_{jr}^{(\beta)}}{\mu_{jr}^{(\beta)}},\
\xi^{(\phi)} = c^{(\phi)} + \sum_{k,h}^{K,H} \frac{\lambda_{kh}^{(\phi)}}{\mu_{kh}^{(\phi)}}. 
\label{update_second}
  \end{equation} 
  
 }
 \caption{Variational inference for binary PMF with covariates.}
 \label{algo_vi}
\end{algorithm}

\subsection{Link prediction}

Given the optimised values of the parameters of the variational approximation $q^\star(\cdot)$ to the posterior, a Monte Carlo posterior model estimate of $\mathbb P(A_{ij}=1)$ can be obtained by averaging \eqref{eq:paij} over $M$ samples from $q^\star(\alpha_{ir}\vert\lambda_{ir}^{(\alpha)},\mu_{ir}^{(\alpha)}),\ q^\star(\beta_{jr}\vert\lambda_{jr}^{(\beta)},\mu_{jr}^{(\beta)})$, and $q^\star(\phi_{kh}\vert\lambda_{kh}^{(\phi)},\mu_{kh}^{(\phi)})$:
\begin{equation}
\hat{\mathbb P}(A_{ij}=1)=1-\frac{1}{M}\nlsum_{m=1}^M \exp\left(-\nlsum_{r=1}^R\alpha_{ir}^{(m)}\beta_{jr}^{(m)}-\nlsum_{h,k}^{K,H}\phi_{kh}^{(m)}x_{ik}y_{jh}\right). \label{full_pval}
\end{equation}

Alternatively, a computationally fast way to approximate ${\mathbb P}(A_{ij}=1)$ plugs in 
the parameters of the estimated variational distributions:
\begin{equation}
\tilde{\mathbb P}(A_{ij}=1|\hat\alpha_{ir}, \hat\beta_{jr}, \hat\phi_{kh})=1-\exp\left(-\nlsum_{r=1}^R\hat\alpha_{ir}\hat{\beta}_{jr}-\nlsum_{h,k}^{K,H}\hat\phi_{kh}x_{ik}y_{jh}\right), \label{biased_pval}
\end{equation}
where, for example,
$\hat\alpha_{ir}=\lambda_{ir}^{(\alpha)}/\mu_{ir}^{(\alpha)}$, the mean of the gamma proxy distribution. 
Note that \eqref{biased_pval} clearly gives a biased estimate, and by Jensen's inequality $\hat{\mathbb P}(A_{ij}=1) \leq \tilde{\mathbb P}(A_{ij}=1)$ in expectation, but it carries a much lower computational burden. The approximation in \eqref{biased_pval} has been successfully used for link prediction and network anomaly detection purposes in \cite{turcotte}.

\section{Seasonal PMF} \label{seasonal_pmf}

The previous sections have been concerned with making inference from a single adjacency matrix $\mvec A$. Now, consider observing a sequence of adjacency matrices $\mvec A_1,\dots,\mvec A_T$, representing snapshots of the same network over time. Further, suppose this time series of adjacency matrices has seasonal dynamics with some known fixed seasonal period, $P$; for example, $P$ could be one day, one week or one year. To recognise time dependence, a third index $t$ is required, such that $A_{ijt}$ denotes the $(i,j)$-th element of the matrix $\mvec A_t$, $t=1,\ldots,T$. 

  As in Section \ref{pmf_covs}, there are assumed to be underlying counts $N_{ijt}$ which are treated as latent variables, and the sequence of observed adjacency matrices is obtained by $A_{ijt}=\mathds 1_{\mathbb N_+}(N_{ijt})$. To account for seasonal repetition in connectivity patterns, the model proposed for the latent counts is:
  \begin{align}
    \label{seasonal_model}
N_{ijt}\sim\ &\mathrm{Pois}\left(\nlsum_{r=1}^R\alpha_{ir}\gamma_{i\gt r}\beta_{jr}\delta_{j\gt r} + \nlsum_{k=1}^{K}\nlsum_{h=1}^H \phi_{kh}
x_{ik}y_{jh} \right) \\
=\ &\mathrm{Pois}\left((\vec\alpha_i\odot\vec\gamma_{i\gt})^\top(\vec\beta_j\odot\vec \delta_{j\gt}) + \vec 1_K^\top(\bm\Phi\odot
\vec x_i\vec y_j^\top)\vec 1_H\right), 
\end{align}
where, for example, $\gt=1+(t\bmod P)$. In general, more complicated functions for $\gt$ might be required, as in Section~\ref{seasonal_sec}.  

The priors on $\alpha_{ir}$ and $\beta_{jr}$ are those given in \eqref{pmf_standard}; these parameters represent a baseline level of activity, which is constant over time. The two additional parameters $\gamma_{i\gt r}$ and $\delta_{j\gt r}$ represent corrections to these rates for seasonal segment $\gt \in \{1,\ldots,P\}$. 
Note that for some applications, it may be anticipated that there is a seasonal adjustment to the rate for the interaction terms of the covariates, in which case temporal adjustments could be also added to $\bm\Phi$.
For identifiability, it is necessary to impose constraints on the seasonal adjustments so that, for example, for all $i,j,r$, $\gamma_{i1r}=\delta_{j1r}=1$.

For $\gt>1$, the following hierarchical priors are placed on $\gamma_{i\gt r}$ and $\delta_{j\gt r}$: 
\begin{gather}
\gamma_{i\gt r}\sim\Gamma(a^{(\gamma)},\zeta_{\gt}^{(\gamma)}),\ \zeta_{\gt}^{(\gamma)}\sim\Gamma(b^{(\gamma)},c^{(\gamma)}),\\ 
\delta_{j\gt r}\sim\Gamma(a^{(\delta)},\zeta_{\gt}^{(\delta)}),\ \zeta_{\gt}^{(\delta)}\sim\Gamma(b^{(\delta)},c^{(\delta)}).
\end{gather} 

Inference for the seasonal model can be performed following the same principles of Section \ref{inference}; full details are given in Appendix~\ref{supp_seasonal}. 
The constraint is implemented in the variational inference framework by setting the variational approximation to $\gamma_{i1r}$ and $\delta_{j1r}$ to a delta function centred at $1$.

\section{Results} \label{results_section}
The extensions to the PMF model detailed in Sections \ref{pmf_covs} and \ref{seasonal_pmf} are used to analyse the LANL authentication data described in Section~\ref{sec:data}. 
In order to assess the predictive performance of the models, the data are split into a training set corresponding to the first 56 days of activity, and a test set corresponding to days 57 through 82. During the latter time period, LANL conducted a \textit{red-team} exercise, where the security team test the robustness of the network by attempting to compromise other network hosts; labels of known compromised authentication events will be used for evaluating the model performance in anomaly detection. 
The parameters are estimated from the training set adjacency matrix, constructed by setting $A_{ij}=1$ if a connection from user $i$ to host $j$ is observed during the training period, and $A_{ij}=0$ otherwise. The predictive performance of the model is then evaluated on the test set adjacency matrix, constructed similarly using the connections observed in the last 25 days. 

\begin{figure}[!b]
\centering
\begin{subfigure}[t]{.495\textwidth}
\caption{\textit{User -- Source }}
\includegraphics[width=\textwidth]{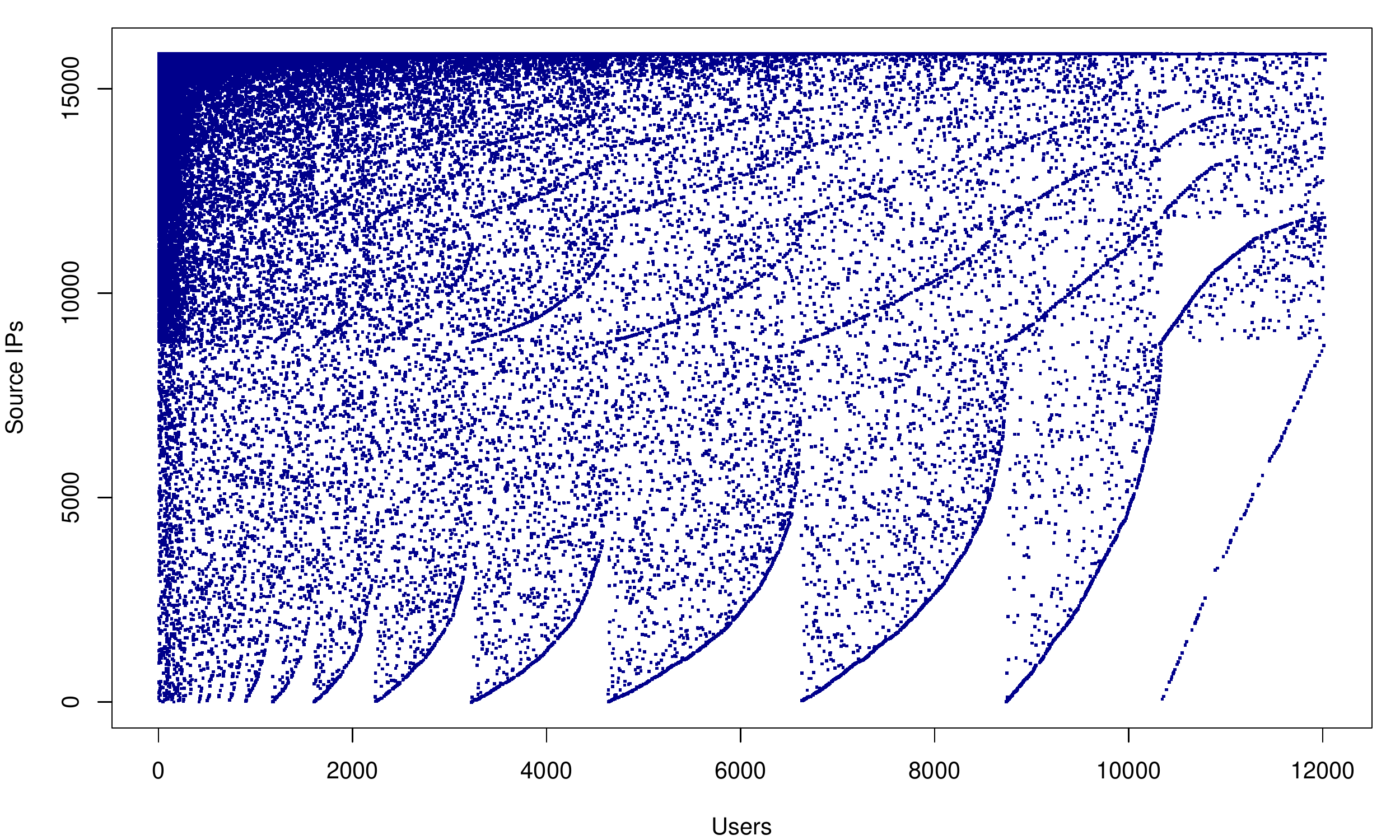}
\label{adjsip}
\end{subfigure}
\begin{subfigure}[t]{.495\textwidth}
\caption{\textit{User -- Destination }}
\includegraphics[width=\textwidth]{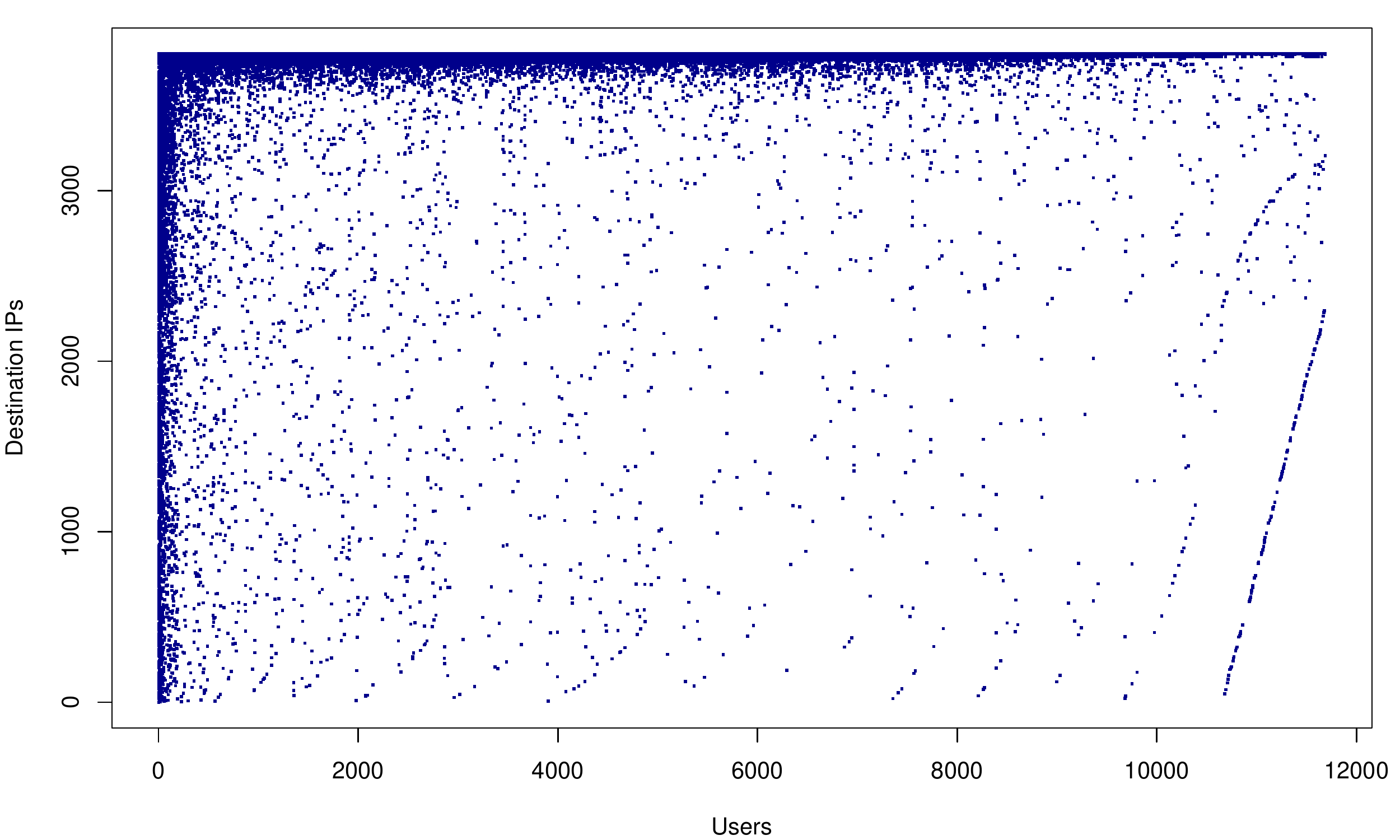}
\label{adjdip}
\end{subfigure}
\caption{Training set adjacency matrices for the two graphs (spy-plot). Nodes are sorted by in-degree and out-degree.}
\label{adj_plots}
\end{figure}

Summary statistics about the data are provided in Table~\ref{summ_table}, where ``cold starts'' refer to links originating from new users and hosts in the test data. Figure~\ref{adj_plots} shows binary heat map plots of the adjacency matrices obtained from the training period for each the two graphs, \textit{User -- Source} and \textit{User -- Destination}.
In all analyses, variational inference is used to estimate the parameters based on the the training data, with a threshold for convergence being $10^{-5}$ for relative difference between two consecutive values of the ELBO \eqref{elbo}.
The prior hyperparameters are set to $a^{*}=b^{*}=1$ and $c^{*}=0.1$, although the algorithm is fairly robust to the choice of these parameters. 
The number of latent features $R=20$ and was chosen using the criterion of the elbow in the scree-plot of singular values. 

 \begin{table}[!b]
\centering
\tabulinesep=1.2mm
\caption{Summary of training and test sets for {\it User -- Destination} and {\it User -- Source}.}
\label{summ_table}
\begin{tabu}{c c c c c}
\cline{2-5}
& \multicolumn{2}{|c|}{{\it User -- Destination }} & \multicolumn{2}{c|}{{\it User -- Source }}  \\
\cline{2-5}
& \multicolumn{1}{|c}{Training set} & \multicolumn{1}{c|}{Test set} & \multicolumn{1}{c}{Training set} & \multicolumn{1}{c|}{Test set} \\
\cline{1-5}
\multicolumn{1}{|c}{Users} & \multicolumn{1}{|c}{$\numprint{11688}$} & \multicolumn{1}{c|}{$\numprint{534}$ new} & \multicolumn{1}{c}{$\numprint{12027}$} & \multicolumn{1}{c|}{$\numprint{507}$ new} \\
\cline{1-5}
\multicolumn{1}{|c}{Hosts} & \multicolumn{1}{|c}{$\numprint{3801}$} & \multicolumn{1}{c|}{$\numprint{1246}$ new} & \multicolumn{1}{c}{$\numprint{15881}$} & \multicolumn{1}{c|}{$\numprint{1236}$ new} \\
\cline{1-5}
\multicolumn{1}{|c}{Links} & \multicolumn{1}{|c}{$\numprint{82517}$} & \multicolumn{1}{c|}{$\numprint{76240}$} & \multicolumn{1}{c}{$\numprint{60059}$} & \multicolumn{1}{c|}{$\numprint{50412}$} \\
\multicolumn{1}{|c}{New links} & \multicolumn{1}{|c}{} & \multicolumn{1}{c|}{$\numprint{11418}$} & \multicolumn{1}{c}{} & \multicolumn{1}{c|}{$\numprint{12080}$} \\
\multicolumn{1}{|c}{Cold starts} & \multicolumn{1}{|c}{} & \multicolumn{1}{c|}{$\numprint{3401}$} & \multicolumn{1}{c}{} & \multicolumn{1}{c|}{$\numprint{3014}$} \\
\cline{1-5}
\end{tabu}
\end{table}

\subsection{Including covariates}

Results are now presented for the extended PMF (EPMF) model, discussed in Section~\ref{pmf_covs}, making use of the nodal covariates described in Section~\ref{sec:data} relating to user roles and categories of computers within the organisation. 
Performance is evaluated by the receiver operating characteristic (ROC) curve and the corresponding area under the curve (AUC). The AUC is used as a measure of quality of classification and will allow for the predictive power of the different models to be ranked. Due to the large computational effort of scoring all entries in the adjacency matrix for EPMF, mostly due to the calculation of $\vec 1_K^\top(\bm\Phi\odot\vec x_i\vec y_j^\top)\vec 1_H$, the AUC is estimated by subsampling the negative class at random from the zeros in the adjacency matrix formed from the test data; 
the sample sizes is chosen to be three times the size of the number of edges in the test set. 
In general, if the sample size is chosen to be at least on the same order of magnitude as $\mathrm{nnz}(\mvec A)$, this procedure leads to reliable estimates of the AUC, as demonstrated via simulation at the end of this subsection. 
The estimated AUC scores are summarised in Table~\ref{auc_table}. For evaluating the AUC scores for new links (edges in the test set not present in the training set), the negative class was also restricted to entries in the training adjacency matrix for which $A_{ij}=0$. 

\begin{table}[!b]
\centering
\tabulinesep=1.2mm
\caption{AUC scores for prediction of \textit{all} and \textit{new links} using standard and extended PMF. Number of latent features: $R=20$.}
\label{auc_table}
\begin{tabu}{c c c c c}
\cline{2-5}
& \multicolumn{2}{|c|}{{\it User -- Destination }} & \multicolumn{2}{c|}{{\it User -- Source }}  \\
\cline{2-5}
& \multicolumn{1}{|c}{PMF} & \multicolumn{1}{c|}{EPMF} & \multicolumn{1}{c}{PMF} & \multicolumn{1}{c|}{EPMF} \\
\cline{1-5}
\multicolumn{1}{|c}{All links} & \multicolumn{1}{|c}{$0.98479$} & \multicolumn{1}{c|}{$0.98707$} & \multicolumn{1}{c}{$0.85797$} & \multicolumn{1}{c|}{$0.96602$} \\
\cline{1-5}
\multicolumn{1}{|c}{New links} & \multicolumn{1}{|c}{$0.95352$} & \multicolumn{1}{c|}{$0.95474$} & \multicolumn{1}{c}{$0.89759$} & \multicolumn{1}{c|}{$0.95260$} \\
\cline{1-5}
\end{tabu}
\end{table}

\begin{figure}[!b]
\centering
\includegraphics[width=.75\textwidth]{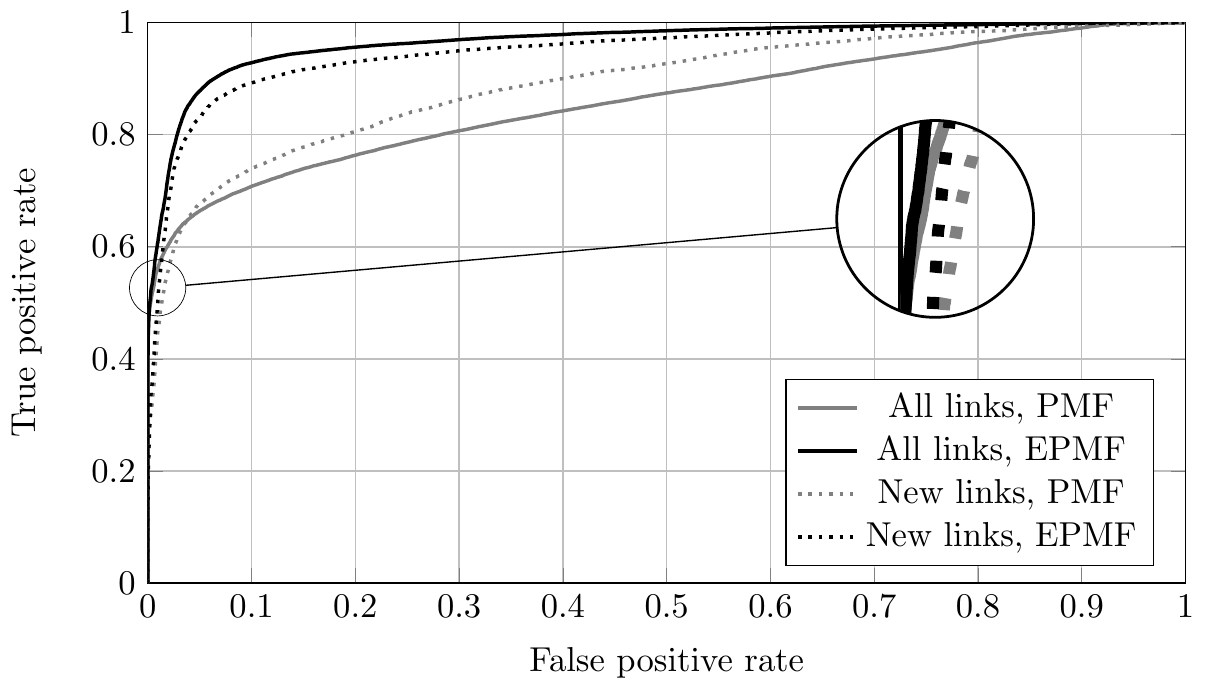}
\caption{ROC curves for standard PMF and extended PMF on \textit{User -- Source}, $R=20$.}
\label{roc}
\end{figure}

Table~\ref{auc_table} shows that the AUC for \textit{User -- Destination} does not change significantly when the extended model is used; however, for \textit{User -- Source}, the extended PMF model offers a significant improvement. The difference in the results between the two networks can be explained by the contrasting structures of the adjacency matrices. The edge density 
for \textit{User -- Destination} is $0.184\%$ and for \textit{User -- Source}, $0.031\%$. However, despite \textit{User -- Destination} having a higher density, the links are concentrated on a small number of dominant nodes, as can be seen in Figure~\ref{adjdip}. Therefore, the prediction task is relatively easy: the probability of a link is roughly approximated by a function of the degree of the node, and adding additional information is not particularly beneficial. For \textit{User -- Source}, as can be seen by Figure~\ref{adjsip}, the links are more evenly distributed between the nodes, and the prediction task is more difficult.
Hence, in this setting, including additional information about known groupings,
for example the location of machines or job category of users, 
is crucial to improve the predictive capability of the model. The ROC curves for the \textit{User -- Source} graph are shown in Figure~\ref{roc}.
 
Some of the covariates might be more predictive than others. To evaluate such differences, the AUC for all links on \textit{User -- Source} have been recalculated excluding each user and host covariate in turn. The difference between the AUC for EPMF fitted with all the covariates, and for EPMF with covariate $k$ removed, could then be used to quantify the predictive power of covariate $k$. According to this methodology, the most relevant user covariates 
are 
\textit{job title}, with a loss in AUC of $0.05442$ when it is removed from the model, followed by 
\textit{location}, with a loss in AUC of $0.03001$. Similarly, for the hosts, the most predictive covariates appear to be 
\textit{location}, with a loss in AUC of $0.05761$, and 
\textit{subnet}, with a loss in AUC of $0.03116$.

In order to assess the accuracy of the AUC approximation obtained from a subsample of the negative class, $500$ simulations have been carried out. 
For standard PMF on \textit{User -- Destination}, estimates of the AUC for all links across different subsampled negative classes had standard deviation $\approx 4.3\cdot10^{-5}$ for a sample size of $3\mathrm{nnz}(\mvec A)$. If $\mathrm{nnz}(\mvec A)$ or $0.1\mathrm{nnz}(\mvec A)$ are used, the standard deviation increases to $\approx 7.6\cdot10^{-5}$ and $\approx 2.4\cdot10^{-4}$ respectively, whereas using $5\mathrm{nnz}(\mvec A)$ gives a value of $\approx 3.5\cdot10^{-5}$.

\subsection{Comparison with Gibbs sampling} \label{gibbs_results}

As discussed in Section~\ref{bayes_inference}, a possible drawback of variational inference is the introduction of bias in the posterior estimates, in particular regarding the variability. Therefore, it is necessary to assess the loss in predictive performance caused by switching to the proposed variational approximation to the posterior. This task is not straightforward, since performing full Bayesian inference on the the standard PMF and EPMF model is computationally extremely demanding when the graph is large, because many posterior samples (usually in the order of tens of thousands) are required to confidently estimate the parameters. 

Appendix~\ref{supp_full} gives details about the conditional distributions required to develop a Gibbs sampler. Table~\ref{gibbs_table} presents the results obtained from $\numprint{10000}$ posterior samples 
with burnin $\numprint{1000}$. The link probabilities have been estimated using the unbiased score \eqref{full_pval}. No issues with convergence of the Markov chain were observed, and the performance seems comparable to the results obtained using variational inference, 
demonstrating that the loss in performance due to the variational approximation seems to be minimal. 

It must be noted that variational inference converges in our application in less than $200$ iterations. For example, for PMF on \textit{User -- Source IP}, the computation time is $\approx 4.5$ seconds per iteration on a MacBook Pro 2017 with a 2.3GHzIntel Core i5 dual-core processor. On the other hand, Gibbs sampling, despite the lower cost of $\approx 2.6$s per iteration, requires many more posterior samples.
Hence, the computational advantages guaranteed by the variational inference procedure are particularly relevant in this context. 

 \begin{table}[!t]
\centering
\tabulinesep=1.2mm
\caption{AUC scores for prediction of \textit{all} and \textit{new links} using Gibbs sampling. Number of latent features: $R=20$.}
\label{gibbs_table}
\begin{tabu}{c c c c c}
\cline{2-5}
& \multicolumn{2}{|c|}{{\it User -- Destination}} & \multicolumn{2}{c|}{{\it User -- Source}}  \\
\cline{2-5}
& \multicolumn{1}{|c}{PMF} & \multicolumn{1}{c|}{EPMF} & \multicolumn{1}{c}{PMF} & \multicolumn{1}{c|}{EPMF} \\
\cline{1-5}
\multicolumn{1}{|c}{All links} & \multicolumn{1}{|c}{$0.98737$} & \multicolumn{1}{c|}{$0.98835$} & \multicolumn{1}{c}{$0.87130$} & \multicolumn{1}{c|}{$0.95809$} \\
\cline{1-5}
\multicolumn{1}{|c}{New links} & \multicolumn{1}{|c}{$0.94311$} & \multicolumn{1}{c|}{$0.94768$} & \multicolumn{1}{c}{$0.83979$} & \multicolumn{1}{c|}{$0.92723$} \\
\cline{1-5}
\end{tabu}
\end{table}

\subsection{Cold starts}

As discussed in Section~\ref{pmf_covs}, the extended PMF model allows for prediction of new entities or nodes in the network (cold starts). To assess performance on links in the test set involving new users or hosts, the estimates of the covariate coefficients $\hat\phi_{kh} = \lambda_{kh}^{(\phi)}/\mu_{kh}^{(\phi)}$ from the training period are used. The latent feature values are set equal to the mean of all users and hosts observed in the training set. 
For comparison against a baseline model, the regular PMF model is used where the latent features are set as above; this has the effect of comparing against the global mean. 

Cold starts can be divided between new \textit{users} and new \textit{hosts}, and the AUC scores for prediction for each case are presented in Table~\ref{cold_starts}. To calculate the AUC, the negative class is randomly sampled from the rows and columns corresponding to the new users and hosts, respectively. Again, there are only minor performance gains for \textit{User -- Destination}, and the regular PMF model using the global average of the latent features provides surprisingly good results. As discussed above, this can be explained by the prediction task being much simpler, and well approximated by a simple degree-based model. In contrast, for the \textit{User -- Source} graph the extended PMF model shows very good predictive performance for cold starts.

 \begin{table}[!t]
\centering
\tabulinesep=1.2mm
\caption{AUC scores for prediction of cold starts.}
\label{cold_starts}
\begin{tabu}{c c c c c}
\cline{2-5}
& \multicolumn{2}{|c|}{{\it User -- Destination }} & \multicolumn{2}{c|}{{\it User -- Source }}  \\
\cline{2-5}
& \multicolumn{1}{|c}{PMF} & \multicolumn{1}{c|}{EPMF} & \multicolumn{1}{c}{PMF} & \multicolumn{1}{c|}{EPMF} \\
\cline{1-5}
\multicolumn{1}{|c}{New Users} & \multicolumn{1}{|c}{$0.96826$} & \multicolumn{1}{c|}{$0.97785$} & \multicolumn{1}{c}{$0.73362$} & \multicolumn{1}{c|}{$0.93148$} \\
\cline{1-5}
\multicolumn{1}{|c}{New Hosts} & \multicolumn{1}{|c}{$0.81789$} & \multicolumn{1}{c|}{$0.82715$} & \multicolumn{1}{c}{$0.79541$} & \multicolumn{1}{c|}{$0.91138$} \\
\cline{1-5}
\end{tabu}
\end{table}

 \subsection{Red-team}
 
\begin{figure}[!b]
\centering
\includegraphics[width=.75\textwidth]{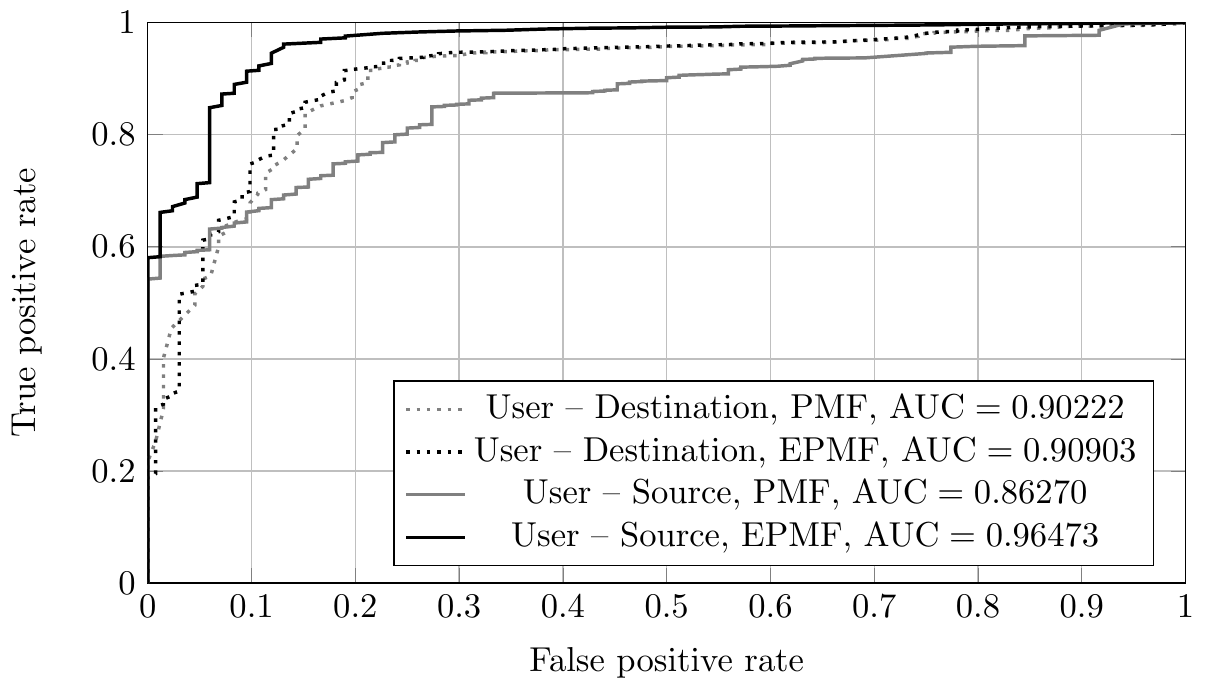}
\caption{ROC curves for prediction of red-team events for standard PMF and extended PMF.}
\label{roc_redteam}
\end{figure}

 The motivation for this work is the detection of cyber attacks; to assess performance from an anomaly detection standpoint, the event labels from the red-team attack are used as a binary classification problem. 
 Figure~\ref{roc_redteam} plots the ROC curves and AUC scores from the standard and extended PMF models, and improvements in detection capability are obtained using EPMF.
Similarly to the previous cases, the predictive performance gain is most notable for \textit{User -- Source}. 

\subsection{Comparisons with alternative prediction methods}

The results in Table~\ref{auc_table} are compared in this section with alternative 
link prediction methods:
\begin{itemize}
\item \textit{Probabilistic matrix factorisation} \citep{prob_mat_fact}: $A_{ij} \sim \bm\alpha_i^\top\bm\beta_j + \varepsilon_{ij},\ \varepsilon_{ij}\overset{iid}{\sim}\mathcal N(0,\sigma^2),$ with $\alpha_{ir}\sim\mathcal N(0,\sigma^2_\alpha)$ and $\beta_{jr}\sim\mathcal N(0,\sigma^2_\beta)$,
where $\mathcal N(\cdot)$ denotes a normal distribution. The parameters were obtained by maximum a posteriori estimation using gradient ascent optimisation techniques. 
\item \textit{Logistic matrix factorisation} \citep{Johnson14}: 
$A_{ij} \sim \mathrm{Bernoulli}(p_{ij})$ where $\mathrm{logit}(p_{ij}) = \bm\alpha_i^\top\bm\beta_j$, with $\alpha_{ir}\sim\mathcal N(0,\sigma^2_\alpha)$ and $\beta_{jr}\sim\mathcal N(0,\sigma^2_\beta)$. 
The parameters were estimated using the same procedure described for probabilistic matrix factorisation.
\item \textit{Bipartite random dot product graph} \citep{Athreya18}: $A_{ij}\sim\mathrm{Bernoulli}\left(\bm\alpha_i^\top\bm\beta_j\right)$.
The latent positions $\bm\alpha_i$ and $\bm\beta_j$ can be estimated using \textit{tSVD} \citep{dhillon}. Assume $\mvec A = \mvec U\mvec D\mvec V^\top + \mvec U_\perp\mvec D_\perp\mvec V^\top_\perp,$
where $\mvec D\in\mathbb R_+^{R\times R}$ is diagonal matrix containing the top $R$ singular values in decreasing order, $\mvec U\in\mathbb R^{\abs{U}\times R}$ and $\mvec V\in\mathbb R^{\abs{V}\times R}$ contain the corresponding left and right singular vectors, and the matrices $\mvec D_\perp$, $\mvec U_\perp$, and $\mvec V_\perp$ contain the remaining singular values and vectors. The tSVD estimates of $\bm\alpha_i$ and $\bm\beta_j$ are respectively the $i$-th and $j$-th row of $\mvec U\mvec D^{1/2}$ and $\mvec V\mvec D^{1/2}$.
A comparison with \textit{tKatz} \citep{dunlavy} is also provided. 
The scores are estimated similarly to tSVD, replacing the diagonal entries of the matrix $\mvec D$ with a transformation of the top $R$ singular values $d_1,\dots,d_R$ of $\mvec A$: $f(d_i)=(1-\eta d_i)^{-1}-1$, with $\eta=10^{-4}$.
\item \textit{Non-negative matrix factorisation} \citep[for example, see][]{Chen17}: nodes are assigned features $\mvec W\in\mathbb R_+^{\abs{U}\times R}$ and $\mvec H\in\mathbb R_+^{\abs{V}\times R}$ obtained as solutions to $\|\mvec A-\mvec W\mvec H^\top\|^2_\text{F}$, where $\|\cdot\|_\text{F}$ is the Frobenius norm. The score associated with any link $(i,j)$ is then obtained as $\vec w_i^\top\vec h_j$, where $\vec w_i^\top$ and $\vec h_j^\top$ are respectively the $i$-th and $j$-th row of $\mvec W$ and $\mvec H$.
\item \textit{Degree-based model}, where the probability of a link is approximated as $\mathbb P(A_{ij}=1)=1-\exp(-d^\mathrm{out}_id^\mathrm{in}_j)$, where $d^\mathrm{out}_i$ and $d^\mathrm{in}_j$ are the out-degree and in-degree of each node.
\end{itemize}


The results are presented in Table~\ref{auc_compare}. Overall, when compared to the results of PMF with Ber-Po link in Table~\ref{auc_table}, the PMF models achieve better results compared to other popular 
techniques, especially for new link prediction. Non-negative matrix factorisation and random dot product graph methods seem to slightly outperform standard PMF when predicting all links, but their performance for new link prediction, a more interesting and difficult task, is significantly worse than PMF.
It must be remarked that some of the alternative methods have been also used for complex applications and purposes other than link prediction. This article does \textit{not} claim that PMF is globally better than such methods, but only that PMF appears to have a better performance for 
link prediction 
on the LANL 
network. 

\begin{table}[!t]
\centering
\tabulinesep=1.2mm
\caption{AUC scores for different link prediction algorithms on the two data sets, with $R=20$.}
\label{auc_compare}
\begin{tabu}{c c c c c}
\cline{2-5}
& \multicolumn{2}{|c|}{{\it User -- Destination }} & \multicolumn{2}{c|}{{\it User -- Source }}  \\
\cline{2-5}
& \multicolumn{1}{|c|}{All links} & \multicolumn{1}{c|}{New links} & \multicolumn{1}{c|}{All links} & \multicolumn{1}{c|}{New links} \\
\cline{1-5}
\multicolumn{1}{|c}{Probabilistic matrix factorisation} & \multicolumn{1}{|c|}{$0.93886$} & \multicolumn{1}{c|}{$0.61006$} & \multicolumn{1}{c|}{$0.83022$} & \multicolumn{1}{c|}{$0.64154$} \\
\cline{1-5}
\multicolumn{1}{|c}{Logistic matrix factorisation} & \multicolumn{1}{|c|}{$0.98079$} & \multicolumn{1}{c|}{$0.95357$} & \multicolumn{1}{c|}{$0.85252$} & \multicolumn{1}{c|}{$0.84330$} \\
\cline{1-5}
\multicolumn{1}{|c}{Random dot product graph -- tSVD} & \multicolumn{1}{|c|}{$0.95050$} & \multicolumn{1}{c|}{$0.69184$} & \multicolumn{1}{c|}{$0.86202$} & \multicolumn{1}{c|}{$0.60935$} \\
\cline{1-5}
\multicolumn{1}{|c}{Random dot product graph -- tKatz} & \multicolumn{1}{|c|}{$0.95392$} & \multicolumn{1}{c|}{$0.71754$} & \multicolumn{1}{c|}{$0.86247$} & \multicolumn{1}{c|}{$0.61120$} \\
\cline{1-5}
\multicolumn{1}{|c}{Non-negative matrix factorisation} & \multicolumn{1}{|c|}{$0.93985$} & \multicolumn{1}{c|}{$0.61675$} & \multicolumn{1}{c|}{$0.86941$} & \multicolumn{1}{c|}{$0.61107$} \\
\cline{1-5}
\multicolumn{1}{|c}{Degree-based model} & \multicolumn{1}{|c|}{$0.95433$} & \multicolumn{1}{c|}{$0.72505$} & \multicolumn{1}{c|}{$0.82719$} & \multicolumn{1}{c|}{$0.70393$} \\
\cline{1-5}
\end{tabu}
\end{table}

In principle, it would be also possible to add covariates to the probabilistic models analysed in this section, including the term $\vec 1_K^\top(\bm\Phi\odot\vec x_i\vec y_j^\top)\vec 1_H$ within the link functions. However, this would be extremely unpractical, since the likelihood for such models is $\mathcal O(\abs{U}\abs{V})$, and the calculation of $\vec x_i\vec y_j^\top$ for all pairs $(i,j)$ is computationally expensive and carries a large memory requirement (18 gigabytes on \textit{User -- Source}), in the order of magnitude of $\mathcal O(\abs{U}\abs{V}KH)$. Therefore, in this section, only the standard models, without the covariate extension, were compared.
On the other hand, fitting EPMF in \eqref{nij} only requires to calculate $\vec x_i\vec y_j^\top$ for all pairs such that $A_{ij}>0$, which is $\mathcal O(\mathrm{nnz}(\mvec A)KH)$. This operation only takes $6.5$ megabytes in memory for \textit{User -- Source}, since $\mvec A$ is extremely sparse in the cyber-security application. This is a significant practical advantage of the proposed model. 

\subsection{Comparison with a joint model} \label{joint_model}

In the previous sections, the graphs \textit{User -- Source} and \textit{User -- Destination} were analysed separately. In this section, the predictive performance of the two individual models is compared to a joint PMF model, where the user-specific latent features are shared between the two graphs. In particular, assume that $\mvec A\in\{0,1\}^{U\times V}$ represents the adjacency matrix for \textit{User -- Source}, and $\mvec A^\prime\in\{0,1\}^{U\times V^\prime}$ for \textit{User -- Destination}. The users are assigned latent positions $\bm\alpha_i,\ i=1,\dots,\abs{U}$ and covariates $\vec x_i$, whereas the source and destination hosts are given latent positions $\bm\beta_j,\ j=1,\dots,\abs{V}$ and $\bm\beta_j^\prime,\ j=1,\dots,\abs{V^\prime}$ respectively, and covariates $\vec y_j$ and $\vec y_j^\prime$. The joint extended PMF model (JEPMF) assumes:
\begin{align}
& A_{ij} = \mathds{1}_{\mathbb N_+}\left(N_{ij}\right),\ N_{ij} \sim \mathrm{Poisson}\left(\bm\alpha_i^\top\bm\beta_j + \vec 1_K^\top(\bm\Phi\odot\vec x_i\vec y_j^\top)\vec 1_H \right), \\ 
& A_{ij}^\prime = \mathds{1}_{\mathbb N_+}\left(N_{ij}^\prime\right),\ N_{ij}^\prime \sim \mathrm{Poisson}\left(\bm\alpha_i^\top\bm\beta_j^\prime + \vec 1_K^\top(\bm\Phi^\prime\odot\vec x_i{\vec y_j^\prime}^\top)\vec 1_{H^\prime}\right).
\label{joint_model}
\end{align}
Similarly, a standard joint PMF model would have the same structure as \eqref{joint_model}, without the covariate term. 
Variational inference for the joint model proceeds similarly to Algorithm~\ref{algo_vi}, with minor differences in the updates for $\lambda_{ir}^{(\alpha)}$ and $\mu_{ir}^{(\alpha)}$. More details are given in Appendix~\ref{supp_joint}. 

The results are presented in Table~\ref{joint_table}. The AUC scores were obtained fitting the joint model 
and then assessing the predictive performance on \textit{User -- Source} and \textit{User -- Destination} separately. Comparing the results with the predictions in Table~\ref{auc_table} for the two individual models, joint PMF produces similar results to the individual models. This suggests that users have a similar behaviour across the two graphs, since adding the constraint of identical user features does not significantly decrease the predictive performance. 
Therefore, it could be concluded that there are latent features for users which vary only slightly between the two graphs, and so are well-suited to a joint modelling approach. This assumption is often made in multiplex networks \citep[see, for example,][]{Kivela14}. 

 \begin{table}[!t]
\centering
\tabulinesep=1.2mm
\caption{AUC scores for prediction of \textit{all} and \textit{new links} using the joint PMF model. Number of latent features: $R=20$.}
\label{joint_table}
\begin{tabu}{c c c c c}
\cline{2-5}
& \multicolumn{2}{|c|}{{\it User -- Destination }} & \multicolumn{2}{c|}{{\it User -- Source }}  \\
\cline{2-5}
& \multicolumn{1}{|c}{PMF} & \multicolumn{1}{c|}{EPMF} & \multicolumn{1}{c}{PMF} & \multicolumn{1}{c|}{EPMF} \\
\cline{1-5}
\multicolumn{1}{|c}{All links} & \multicolumn{1}{|c}{$0.98206$} & \multicolumn{1}{c|}{$0.98290$} & \multicolumn{1}{c}{$0.86061$} & \multicolumn{1}{c|}{$0.95585$} \\
\cline{1-5}
\multicolumn{1}{|c}{New links} & \multicolumn{1}{|c}{$0.94563$} & \multicolumn{1}{c|}{$0.94565$} & \multicolumn{1}{c}{$0.89414$} & \multicolumn{1}{c|}{$0.94637$} \\
\cline{1-5}
\end{tabu}
\end{table}

\subsection{Seasonal modelling} \label{seasonal_sec}
To investigate dynamic modelling, binary adjacency matrices $\mvec A_1,\dots,\mvec A_{82}$ are calculated for each day across the train and test periods. The seasonal PMF model with the inclusion of covariates (SEPMF) \eqref{seasonal_model} is then compared against EPMF; for EPMF, the adjacency matrices are assumed to be independent realisations randomly generated from a fixed set of latent features. Due to a ``9 day-80 hour'' work schedule operated at LANL, whereby employees can elect to take vacation every other Friday, the seasonal period is assumed to be comprised of four segments: weekdays (Monday - Thursday), weekends (Saturday and Sunday), and two separate segments for alternating Friday's. For each model, binary classification is performed using the model predictive scores calculated across the entire period. For the positive class, scores are calculated for all user-host pairs $(i,j)$ such that $A_{ijt}=1$ for at least one $t$ in the test set; for the negative class, a random sample of $(i,j)$ pairs such that $A_{ijt}=0$ for all $t$ in the test set are obtained, with sample size equal to three times the total number of observed links.

Table~\ref{auc_seasonal} presents the resulting AUC scores. For both networks, the seasonal model does not globally outperform the extended PMF model for {\it all links}. However, improvements are obtained for prediction of the \textit{new links}. One explanation for the weaker overall performance could be the reduced training sample size implied for the seasonal model: EPMF in a dynamic setting assumes that the all daily graphs have been sampled from the same process, whereas if the seasonal model is used then the daily graphs are only informative for the corresponding seasonal segments. In addition, as briefly mentioned in Section \ref{intro}, elements of the data exhibit strong polling patterns, often due to computers automatically authenticating on users' behalves \citep{Turcotte18}; some of the links that exhibit polling will not exhibit seasonal patterns, as the human behaviour has not been separated from the automated behaviour.

 \begin{table}[!t]
\centering
\tabulinesep=1.2mm
\caption{AUC scores for prediction of \textit{all} and \textit{new links} using seasonal PMF.}
\label{auc_seasonal}
\begin{tabu}{c c c c c}
\cline{2-5}
& \multicolumn{2}{|c|}{{\it User -- Destination }} & \multicolumn{2}{c|}{{\it User -- Source }}  \\
\cline{2-5}
& \multicolumn{1}{|c}{EPMF} & \multicolumn{1}{c|}{SEPMF} & \multicolumn{1}{c}{EPMF} & \multicolumn{1}{c|}{SEPMF} \\
\cline{1-5}
\multicolumn{1}{|c}{All links} & \multicolumn{1}{|c}{$0.96550$} & \multicolumn{1}{c|}{$0.96205$} & \multicolumn{1}{c}{$0.93559$} & \multicolumn{1}{c|}{$0.92829$} \\
\cline{1-5}
\multicolumn{1}{|c}{New links} & \multicolumn{1}{|c}{$0.87107$} & \multicolumn{1}{c|}{$0.89337$} & \multicolumn{1}{c}{$0.85009$} & \multicolumn{1}{c|}{$0.85748$} \\
\cline{1-5}
\end{tabu}
\end{table}

 On the other hand, improvements in the estimation of new links, despite the reduced training sample size, demonstrates that it can be beneficial to understand the temporal dynamics of the network for these cases. Considering the context of the application, it might be perfectly normal for a user to authenticate to a computer during the week; however, that same authentication would be extremely unusual on the weekends when the user is not present at work. Without the seasonal model this behavioural difference in would be missed. 
 
\section{Conclusion and discussion}

Extensions of the standard Poisson matrix factorisation model have been proposed, motivated by applications to computer network data, in particular the LANL enterprise computer network. 
The extensions are threefold: handling binary matrices, including covariates for users and hosts in the PMF framework, and accounting for seasonal effects.
The counts $N_{ijt}$ have been treated as censored, and it has been assumed that only the binary indicator $A_{ijt}=\mathds 1_{\mathbb N_+}(N_{ijt})$ is observed. Starting from the hierarchical Poisson matrix factorisation model of \cite{gopalan}, which only includes the latent features $\bm\alpha_i$ and $\bm\beta_j$, covariates have been included through the matrix of coefficients $\bm\Phi$. Seasonal adjustments for the coefficients are obtained through the variables $\bm\gamma_{i\gt}$ and $\bm\delta_{j\gt}$. 
A variational inference algorithm is proposed, suitably adapted for the Bernoulli-Poisson link. This article mainly considered categorical covariates, but the methodology could be extended to include other forms of covariates, with minimal modifications to the inferential algorithms.

The results show improvements over alternative models for link prediction purposes on the real computer network data. Including covariates provides significant uplift in predictive performance and allows prediction for new nodes arriving into the network. 
This discovery provides valuable insight on the potential benefits of including nodal covariates when inferring network structure. In particular, user covariates like job title and location were found to be particularly informative for the users, and location and subnet for the hosts. 
The seasonal model provides time-varying anomaly scores and offers marginal improvements for predicting new links, which are of primary interest in cyber-security applications.

Despite the focus on bipartite graphs here, the proposed methodologies could be readily adapted to undirected and general directed graphs. For an undirected graph, the PMF model with Ber-Po link would assume:
\begin{equation}
 A_{ij} = \mathds{1}_{\mathbb N_+}\left(N_{ij}\right),\ N_{ij} \sim \mathrm{Poisson}\left(\bm\alpha_i^\top\bm\alpha_j\right),\ i<j,\ A_{ij} = A_{ji}. \label{und_graph}
\end{equation}
For directed graphs on the same node set, it could be assumed that each node has the same behaviour as source and destination of the connection, implying equation \eqref{und_graph} for undirected graphs applies, removing the constraint $A_{ij} = A_{ji}$. Alternatively, each node could be given two latent features: $\bm\alpha_i$ for its behaviour as source and $\bm\beta_i$ for its behaviour as destination. This is a special case of the bipartite graph model, where $U\equiv V$, hence \eqref{nij} applies. Note that mean-field variational inference within the directed graph context might be problematic, since the mean-field assumption $\mathrm{Cov}(N_{ij},N_{ji})=0$ would be unrealistic if network reciprocity is observed. This is often the case in cyber-security applications, for example in network flow data, representing directed summaries of connections between IP addresses.


\section*{Acknowledgements}
The authors acknowledge funding from Los Alamos National Laboratory, EPSRC and the Heilbronn Institute for Mathematical Research.
Research presented in this article was supported by the Laboratory Directed Research and Development program of Los Alamos National Laboratory (New Mexico, USA) under project number 20180607ECR. 

\bibliography{biblio}

\begin{thebibliography}{59}

\bibitem[\protect\citeauthoryear{Acharya et~al.}{2015}]{acharya}
\begin{binproceedings}[author]
\bauthor{\bsnm{Acharya},~\bfnm{A.}\binits{A.}},
  \bauthor{\bsnm{Teffer},~\bfnm{D.}\binits{D.}},
  \bauthor{\bsnm{Henderson},~\bfnm{J.}\binits{J.}},
  \bauthor{\bsnm{Tyler},~\bfnm{M.}\binits{M.}},
  \bauthor{\bsnm{Zhou},~\bfnm{M.}\binits{M.}} \AND
  \bauthor{\bsnm{Ghosh},~\bfnm{J.}\binits{J.}}
(\byear{2015}).
\btitle{Gamma {P}rocess {P}oisson Factorization for Joint Modeling of Network
  and Documents}.
In \bbooktitle{Proceedings of the European Conference on Machine Learning and
  Knowledge Discovery in Databases}
\bvolume{1}
\bpages{283--299}.
\end{binproceedings}
\endbibitem

\bibitem[\protect\citeauthoryear{Adomavicius and Tuzhilin}{2005}]{adomavicius}
\begin{barticle}[author]
\bauthor{\bsnm{Adomavicius},~\bfnm{G.}\binits{G.}} \AND
  \bauthor{\bsnm{Tuzhilin},~\bfnm{A.}\binits{A.}}
(\byear{2005}).
\btitle{Toward the Next Generation of Recommender Systems: A Survey of the
  State-of-the-Art and Possible Extensions}.
\bjournal{IEEE Transactions on Knowledge and Data Engineering}
\bvolume{17}
\bpages{734--749}.
\end{barticle}
\endbibitem

\bibitem[\protect\citeauthoryear{Agarwal, Zhang and
  Mazumder}{2011}]{agarwal_aoas}
\begin{barticle}[author]
\bauthor{\bsnm{Agarwal},~\bfnm{D.}\binits{D.}},
  \bauthor{\bsnm{Zhang},~\bfnm{L.}\binits{L.}} \AND
  \bauthor{\bsnm{Mazumder},~\bfnm{R.}\binits{R.}}
(\byear{2011}).
\btitle{Modeling item-item similarities for personalized recommendations on
  {Y}ahoo! front page}.
\bjournal{Annals of Applied Statistics}
\bvolume{5}
\bpages{1839--1875}.
\end{barticle}
\endbibitem

\bibitem[\protect\citeauthoryear{Amit et~al.}{2019}]{Amit19}
\begin{binproceedings}[author]
\bauthor{\bsnm{Amit},~\bfnm{Idan}\binits{I.}},
  \bauthor{\bsnm{Matherly},~\bfnm{John}\binits{J.}},
  \bauthor{\bsnm{Hewlett},~\bfnm{William}\binits{W.}},
  \bauthor{\bsnm{Xu},~\bfnm{Zhi}\binits{Z.}},
  \bauthor{\bsnm{Meshi},~\bfnm{Yinnon}\binits{Y.}} \AND
  \bauthor{\bsnm{Weinberger},~\bfnm{Yigal}\binits{Y.}}
(\byear{2019}).
\btitle{{Machine Learning in Cyber-Security - Problems, Challenges and Data
  Sets}}.
In \bbooktitle{AAAI-19 Workshop on Engineering Dependable and Secure Machine
  Learning Systems}.
\end{binproceedings}
\endbibitem

\bibitem[\protect\citeauthoryear{Anderson et~al.}{2018}]{Anderson18}
\begin{binbook}[author]
\bauthor{\bsnm{Anderson},~\bfnm{Blake}\binits{B.}},
  \bauthor{\bsnm{Vejman},~\bfnm{Martin}\binits{M.}},
  \bauthor{\bsnm{McGrew},~\bfnm{David}\binits{D.}} \AND
  \bauthor{\bsnm{Paul},~\bfnm{Subharthi}\binits{S.}}
(\byear{2018}).
\btitle{Towards Generalisable Network Threat Detection}
In \bbooktitle{Data Science for Cyber-Security}
\bchapter{4},
\bpages{77-94}.
\bpublisher{World Scientific}.
\end{binbook}
\endbibitem

\bibitem[\protect\citeauthoryear{Athreya et~al.}{2018}]{Athreya18}
\begin{barticle}[author]
\bauthor{\bsnm{Athreya},~\bfnm{Avanti}\binits{A.}},
  \bauthor{\bsnm{Fishkind},~\bfnm{Donniell~E.}\binits{D.~E.}},
  \bauthor{\bsnm{Tang},~\bfnm{Minh}\binits{M.}},
  \bauthor{\bsnm{Priebe},~\bfnm{Carey~E.}\binits{C.~E.}},
  \bauthor{\bsnm{Park},~\bfnm{Youngser}\binits{Y.}},
  \bauthor{\bsnm{Vogelstein},~\bfnm{Joshua~T.}\binits{J.~T.}},
  \bauthor{\bsnm{Levin},~\bfnm{Keith}\binits{K.}},
  \bauthor{\bsnm{Lyzinski},~\bfnm{Vince}\binits{V.}},
  \bauthor{\bsnm{Qin},~\bfnm{Yichen}\binits{Y.}} \AND
  \bauthor{\bsnm{Sussman},~\bfnm{Daniel~L}\binits{D.~L.}}
(\byear{2018}).
\btitle{Statistical Inference on Random Dot Product Graphs: a Survey}.
\bjournal{Journal of Machine Learning Research}
\bvolume{18}
\bpages{1-92}.
\end{barticle}
\endbibitem

\bibitem[\protect\citeauthoryear{Bishop}{2006}]{bishop}
\begin{bbook}[author]
\bauthor{\bsnm{Bishop},~\bfnm{C.~M.}\binits{C.~M.}}
(\byear{2006}).
\btitle{Pattern Recognition and Machine Learning}.
\bseries{Information Science and Statistics}.
\bpublisher{Springer-Verlag}, \baddress{Berlin, Heidelberg}.
\end{bbook}
\endbibitem

\bibitem[\protect\citeauthoryear{Blei, Kucukelbir and McAuliffe}{2017}]{blei}
\begin{barticle}[author]
\bauthor{\bsnm{Blei},~\bfnm{D.~M.}\binits{D.~M.}},
  \bauthor{\bsnm{Kucukelbir},~\bfnm{A.}\binits{A.}} \AND
  \bauthor{\bsnm{McAuliffe},~\bfnm{J.~D.}\binits{J.~D.}}
(\byear{2017}).
\btitle{Variational Inference: A Review for Statisticians}.
\bjournal{Journal of the American Statistical Association}
\bvolume{112}
\bpages{859-877}.
\end{barticle}
\endbibitem

\bibitem[\protect\citeauthoryear{Canny}{2004}]{canny}
\begin{binproceedings}[author]
\bauthor{\bsnm{Canny},~\bfnm{J.}\binits{J.}}
(\byear{2004}).
\btitle{{GaP}: A Factor Model for Discrete Data}.
In \bbooktitle{Proceedings of the 27th Annual International ACM SIGIR
  Conference}.
\bseries{SIGIR '04}
\bpages{122--129}.
\end{binproceedings}
\endbibitem

\bibitem[\protect\citeauthoryear{Cemgil}{2009}]{cemgil}
\begin{barticle}[author]
\bauthor{\bsnm{Cemgil},~\bfnm{A.~T.}\binits{A.~T.}}
(\byear{2009}).
\btitle{Bayesian Inference for Nonnegative Matrix Factorisation Models}.
\bjournal{Computational Intelligence and Neuroscience}.
\end{barticle}
\endbibitem

\bibitem[\protect\citeauthoryear{Chaney, Blei and Eliassi-Rad}{2015}]{chaney}
\begin{binproceedings}[author]
\bauthor{\bsnm{Chaney},~\bfnm{A.~J.~B.}\binits{A.~J.~B.}},
  \bauthor{\bsnm{Blei},~\bfnm{D.~M.}\binits{D.~M.}} \AND
  \bauthor{\bsnm{Eliassi-Rad},~\bfnm{T.}\binits{T.}}
(\byear{2015}).
\btitle{A Probabilistic Model for Using Social Networks in Personalized Item
  Recommendation}.
In \bbooktitle{Proceedings of the 9th ACM Conference on Recommender Systems}.
\bseries{RecSys '15}
\bpages{43--50}.
\bpublisher{ACM}.
\end{binproceedings}
\endbibitem

\bibitem[\protect\citeauthoryear{Charlin et~al.}{2015}]{charlin}
\begin{binproceedings}[author]
\bauthor{\bsnm{Charlin},~\bfnm{L.}\binits{L.}},
  \bauthor{\bsnm{Ranganath},~\bfnm{R.}\binits{R.}},
  \bauthor{\bsnm{McInerney},~\bfnm{J.}\binits{J.}} \AND
  \bauthor{\bsnm{Blei},~\bfnm{D.~M.}\binits{D.~M.}}
(\byear{2015}).
\btitle{Dynamic {P}oisson Factorization}.
In \bbooktitle{Proceedings of the 9th ACM Conference on Recommender Systems}
\bpages{155--162}.
\bpublisher{ACM}.
\end{binproceedings}
\endbibitem

\bibitem[\protect\citeauthoryear{Chen et~al.}{2017}]{Chen17}
\begin{barticle}[author]
\bauthor{\bsnm{Chen},~\bfnm{Bolun}\binits{B.}},
  \bauthor{\bsnm{Li},~\bfnm{Fenfen}\binits{F.}},
  \bauthor{\bsnm{Chen},~\bfnm{Senbo}\binits{S.}},
  \bauthor{\bsnm{Hu},~\bfnm{Ronglin}\binits{R.}} \AND
  \bauthor{\bsnm{Chen},~\bfnm{Ling}\binits{L.}}
(\byear{2017}).
\btitle{Link prediction based on non-negative matrix factorization}.
\bjournal{PLOS ONE}
\bvolume{12}
\bpages{1-18}.
\end{barticle}
\endbibitem

\bibitem[\protect\citeauthoryear{Clauset, Moore and Newman}{2008}]{Clauset08}
\begin{barticle}[author]
\bauthor{\bsnm{Clauset},~\bfnm{Aaron}\binits{A.}},
  \bauthor{\bsnm{Moore},~\bfnm{Cristopher}\binits{C.}} \AND
  \bauthor{\bsnm{Newman},~\bfnm{M.~E.~J.}\binits{M.~E.~J.}}
(\byear{2008}).
\btitle{Hierarchical structure and the prediction of missing links in
  networks}.
\bjournal{Nature}
\bvolume{453}.
\end{barticle}
\endbibitem

\bibitem[\protect\citeauthoryear{da~Silva, Langseth and
  Ramampiaro}{2017}]{da_silva}
\begin{binproceedings}[author]
\bauthor{\bparticle{da} \bsnm{Silva},~\bfnm{E.~d.~S.}\binits{E.~d.~S.}},
  \bauthor{\bsnm{Langseth},~\bfnm{H.}\binits{H.}} \AND
  \bauthor{\bsnm{Ramampiaro},~\bfnm{H.}\binits{H.}}
(\byear{2017}).
\btitle{Content-Based Social Recommendation with {P}oisson Matrix
  Factorization}.
In \bbooktitle{Machine Learning and Knowledge Discovery in Databases}
\bpages{530--546}.
\end{binproceedings}
\endbibitem

\bibitem[\protect\citeauthoryear{Dai et~al.}{2019}]{Dai19}
\begin{barticle}[author]
\bauthor{\bsnm{Dai},~\bfnm{Ben}\binits{B.}},
  \bauthor{\bsnm{Wang},~\bfnm{Junhui}\binits{J.}},
  \bauthor{\bsnm{Shen},~\bfnm{Xiaotong}\binits{X.}} \AND
  \bauthor{\bsnm{Qu},~\bfnm{Annie}\binits{A.}}
(\byear{2019}).
\btitle{Smooth neighborhood recommender systems}.
\bjournal{Journal of Machine Learning Research}
\bvolume{20}
\bpages{1-24}.
\end{barticle}
\endbibitem

\bibitem[\protect\citeauthoryear{Dhillon}{2001}]{dhillon}
\begin{binproceedings}[author]
\bauthor{\bsnm{Dhillon},~\bfnm{I.~S.}\binits{I.~S.}}
(\byear{2001}).
\btitle{Co-clustering Documents and Words Using Bipartite Spectral Graph
  Partitioning}.
In \bbooktitle{Proceedings of the Seventh ACM SIGKDD Conference on Knowledge
  Discovery and Data Mining}.
\bseries{KDD '01}
\bpages{269--274}.
\bpublisher{ACM}, \baddress{New York, NY, USA}.
\end{binproceedings}
\endbibitem

\bibitem[\protect\citeauthoryear{Dunlavy, Kolda and Acar}{2011}]{dunlavy}
\begin{barticle}[author]
\bauthor{\bsnm{Dunlavy},~\bfnm{D.~M.}\binits{D.~M.}},
  \bauthor{\bsnm{Kolda},~\bfnm{T.~G.}\binits{T.~G.}} \AND
  \bauthor{\bsnm{Acar},~\bfnm{E.}\binits{E.}}
(\byear{2011}).
\btitle{Temporal link prediction using matrix and tensor factorizations}.
\bjournal{ACM Transactions on Knowledge Discovery from Data}
\bvolume{5}.
\end{barticle}
\endbibitem

\bibitem[\protect\citeauthoryear{Dunson and Herring}{2005}]{dunson}
\begin{barticle}[author]
\bauthor{\bsnm{Dunson},~\bfnm{D.~B.}\binits{D.~B.}} \AND
  \bauthor{\bsnm{Herring},~\bfnm{A.~H.}\binits{A.~H.}}
(\byear{2005}).
\btitle{{B}ayesian latent variable models for mixed discrete outcomes}.
\bjournal{Biostatistics}
\bvolume{6}
\bpages{11-25}.
\end{barticle}
\endbibitem

\bibitem[\protect\citeauthoryear{Fienberg}{2012}]{Fienberg12}
\begin{barticle}[author]
\bauthor{\bsnm{Fienberg},~\bfnm{Stephen~E.}\binits{S.~E.}}
(\byear{2012}).
\btitle{A Brief History of Statistical Models for Network Analysis and Open
  Challenges}.
\bjournal{Journal of Computational and Graphical Statistics}
\bvolume{21}
\bpages{825-839}.
\end{barticle}
\endbibitem

\bibitem[\protect\citeauthoryear{Fithian and Mazumder}{2018}]{Fithian18}
\begin{barticle}[author]
\bauthor{\bsnm{Fithian},~\bfnm{William}\binits{W.}} \AND
  \bauthor{\bsnm{Mazumder},~\bfnm{Rahul}\binits{R.}}
(\byear{2018}).
\btitle{Flexible Low-Rank Statistical Modeling with Side Information}.
\bjournal{Statistical Science}
\bvolume{33}
\bpages{238--260}.
\end{barticle}
\endbibitem

\bibitem[\protect\citeauthoryear{Goldenberg et~al.}{2010}]{Goldenberg10}
\begin{barticle}[author]
\bauthor{\bsnm{Goldenberg},~\bfnm{Anna}\binits{A.}},
  \bauthor{\bsnm{Zheng},~\bfnm{Alice~X.}\binits{A.~X.}},
  \bauthor{\bsnm{Fienberg},~\bfnm{Stephen~E.}\binits{S.~E.}} \AND
  \bauthor{\bsnm{Airoldi},~\bfnm{Edoardo~M.}\binits{E.~M.}}
(\byear{2010}).
\btitle{A Survey of Statistical Network Models}.
\bjournal{Found. Trends Mach. Learn.}
\bvolume{2}
\bpages{129--233}.
\end{barticle}
\endbibitem

\bibitem[\protect\citeauthoryear{Gopalan, Charlin and Blei}{2014}]{gopalan_cpf}
\begin{binproceedings}[author]
\bauthor{\bsnm{Gopalan},~\bfnm{P.}\binits{P.}},
  \bauthor{\bsnm{Charlin},~\bfnm{L.}\binits{L.}} \AND
  \bauthor{\bsnm{Blei},~\bfnm{D.~M.}\binits{D.~M.}}
(\byear{2014}).
\btitle{Content-based Recommendations with {P}oisson Factorization}.
In \bbooktitle{Proceedings of the 27th International Conference on Neural
  Information Processing Systems}.
\bseries{NIPS'14}
\bvolume{2}
\bpages{3176--3184}.
\bpublisher{MIT Press}.
\end{binproceedings}
\endbibitem

\bibitem[\protect\citeauthoryear{Gopalan, Hofman and Blei}{2015}]{gopalan}
\begin{binproceedings}[author]
\bauthor{\bsnm{Gopalan},~\bfnm{P.}\binits{P.}},
  \bauthor{\bsnm{Hofman},~\bfnm{J.~M.}\binits{J.~M.}} \AND
  \bauthor{\bsnm{Blei},~\bfnm{D.~M.}\binits{D.~M.}}
(\byear{2015}).
\btitle{Scalable Recommendation with Hierarchical {P}oisson Factorization}.
In \bbooktitle{Proceedings of the 31st Conference on Uncertainty in Artificial
  Intelligence}.
\bseries{UAI'15}
\bpages{326--335}.
\bpublisher{AUAI Press}.
\end{binproceedings}
\endbibitem

\bibitem[\protect\citeauthoryear{Heard, Rubin-Delanchy and
  Lawson}{2014}]{Heard14}
\begin{barticle}[author]
\bauthor{\bsnm{Heard},~\bfnm{N.~A.}\binits{N.~A.}},
  \bauthor{\bsnm{Rubin-Delanchy},~\bfnm{P.~T.~G.}\binits{P.~T.~G.}} \AND
  \bauthor{\bsnm{Lawson},~\bfnm{D.~J.}\binits{D.~J.}}
(\byear{2014}).
\btitle{Filtering automated polling traffic in computer network flow data}.
\bjournal{Proceedings - 2014 IEEE Joint Intelligence and Security Informatics
  Conference, JISIC 2014}
\bpages{268-271}.
\end{barticle}
\endbibitem

\bibitem[\protect\citeauthoryear{Heard et~al.}{2018}]{HeardAdams18}
\begin{bbook}[author]
\bauthor{\bsnm{Heard},~\bfnm{N.~A.}\binits{N.~A.}},
  \bauthor{\bsnm{Adams},~\bfnm{N.}\binits{N.}},
  \bauthor{\bsnm{Rubin-Delanchy},~\bfnm{P.}\binits{P.}} \AND
  \bauthor{\bsnm{Turcotte},~\bfnm{M.}\binits{M.}}
(\byear{2018}).
\btitle{Data Science for Cyber-Security}.
\bpublisher{World Scientific (Europe)}.
\end{bbook}
\endbibitem

\bibitem[\protect\citeauthoryear{Hern\'{a}ndez-Lobato, Houlsby and
  Ghahramani}{2014}]{Lobato14}
\begin{binproceedings}[author]
\bauthor{\bsnm{Hern\'{a}ndez-Lobato},~\bfnm{J.~M.}\binits{J.~M.}},
  \bauthor{\bsnm{Houlsby},~\bfnm{N.}\binits{N.}} \AND
  \bauthor{\bsnm{Ghahramani},~\bfnm{Z.}\binits{Z.}}
(\byear{2014}).
\btitle{Stochastic Inference for Scalable Probabilistic Modeling of Binary
  Matrices}.
In \bbooktitle{Proceedings of the 31st International Conference on Machine
  Learning}.
\bseries{ICML'14}
\bvolume{32}
\bpages{II-379--II-387}.
\end{binproceedings}
\endbibitem

\bibitem[\protect\citeauthoryear{Hoff}{2005}]{hoff_bilinear}
\begin{barticle}[author]
\bauthor{\bsnm{Hoff},~\bfnm{P.~D.}\binits{P.~D.}}
(\byear{2005}).
\btitle{Bilinear Mixed-Effects Models for Dyadic Data}.
\bjournal{Journal of the American Statistical Association}
\bvolume{100}
\bpages{286-295}.
\end{barticle}
\endbibitem

\bibitem[\protect\citeauthoryear{Hoff, Raftery and Handcock}{2002}]{hoff}
\begin{barticle}[author]
\bauthor{\bsnm{Hoff},~\bfnm{P.~D}\binits{P.~D.}},
  \bauthor{\bsnm{Raftery},~\bfnm{A.~E.}\binits{A.~E.}} \AND
  \bauthor{\bsnm{Handcock},~\bfnm{M.~S.}\binits{M.~S.}}
(\byear{2002}).
\btitle{Latent space approaches to social network analysis}.
\bjournal{Journal of the American Statistical Association}
\bvolume{97}
\bpages{1090--1098}.
\end{barticle}
\endbibitem

\bibitem[\protect\citeauthoryear{Hosseini et~al.}{2018}]{Hosseini18}
\begin{barticle}[author]
\bauthor{\bsnm{Hosseini},~\bfnm{S.}\binits{S.}},
  \bauthor{\bsnm{Khodadadi},~\bfnm{A.}\binits{A.}},
  \bauthor{\bsnm{Alizadeh},~\bfnm{K.}\binits{K.}},
  \bauthor{\bsnm{Arabzadeh},~\bfnm{A.}\binits{A.}},
  \bauthor{\bsnm{Farajtabar},~\bfnm{M.}\binits{M.}},
  \bauthor{\bsnm{Zha},~\bfnm{H.}\binits{H.}} \AND
  \bauthor{\bsnm{Rabiee},~\bfnm{H.~R.~R.}\binits{H.~R.~R.}}
(\byear{2018}).
\btitle{Recurrent {P}oisson factorization for temporal recommendation}.
\bjournal{IEEE Transactions on Knowledge and Data Engineering}.
\end{barticle}
\endbibitem

\bibitem[\protect\citeauthoryear{Huggins et~al.}{2019}]{Huggins19}
\begin{binproceedings}[author]
\bauthor{\bsnm{Huggins},~\bfnm{Jonathan~H.}\binits{J.~H.}},
  \bauthor{\bsnm{Campbell},~\bfnm{Trevor}\binits{T.}},
  \bauthor{\bsnm{Kasprzak},~\bfnm{Mikolaj}\binits{M.}} \AND
  \bauthor{\bsnm{Broderick},~\bfnm{Tamara}\binits{T.}}
(\byear{2019}).
\btitle{Scalable Gaussian Process Inference with Finite-data Mean and Variance
  Guarantees}.
In \bbooktitle{Proceedings of Machine Learning Research}
\bvolume{89}
\bpages{796--805}.
\end{binproceedings}
\endbibitem

\bibitem[\protect\citeauthoryear{Jeske et~al.}{2018}]{Jeske18}
\begin{barticle}[author]
\bauthor{\bsnm{Jeske},~\bfnm{D.~R.}\binits{D.~R.}},
  \bauthor{\bsnm{Stevens},~\bfnm{N.~T.}\binits{N.~T.}},
  \bauthor{\bsnm{Tartakovsky},~\bfnm{A.~G.}\binits{A.~G.}} \AND
  \bauthor{\bsnm{Wilson},~\bfnm{J.~D.}\binits{J.~D.}}
(\byear{2018}).
\btitle{Statistical methods for network surveillance}.
\bjournal{Applied Stochastic Models in Business and Industry}
\bvolume{34}
\bpages{425-445}.
\end{barticle}
\endbibitem

\bibitem[\protect\citeauthoryear{Johnson}{2014}]{Johnson14}
\begin{binproceedings}[author]
\bauthor{\bsnm{Johnson},~\bfnm{C.~C}\binits{C.~C.}}
(\byear{2014}).
\btitle{Logistic matrix factorization for implicit feedback data}.
In \bbooktitle{Proceedings of the NIPS 2014 Workshop on Distributed Machine
  Learning and Matrix Computations}.
\end{binproceedings}
\endbibitem

\bibitem[\protect\citeauthoryear{Khanna et~al.}{2013}]{Khanna13}
\begin{binproceedings}[author]
\bauthor{\bsnm{Khanna},~\bfnm{R.}\binits{R.}},
  \bauthor{\bsnm{Zhang},~\bfnm{L.}\binits{L.}},
  \bauthor{\bsnm{Agarwal},~\bfnm{D.}\binits{D.}} \AND
  \bauthor{\bsnm{Chen},~\bfnm{B.~C.}\binits{B.~C.}}
(\byear{2013}).
\btitle{Parallel matrix factorization for binary response}.
In \bbooktitle{IEEE International Conference on Big Data 2013}
\bpages{430--438}.
\end{binproceedings}
\endbibitem

\bibitem[\protect\citeauthoryear{{Kim} et~al.}{2017}]{Kim17}
\begin{barticle}[author]
\bauthor{\bsnm{{Kim}},~\bfnm{Bomin}\binits{B.}},
  \bauthor{\bsnm{{Lee}},~\bfnm{Kevin}\binits{K.}},
  \bauthor{\bsnm{{Xue}},~\bfnm{Lingzhou}\binits{L.}} \AND
  \bauthor{\bsnm{{Niu}},~\bfnm{Xiaoyue}\binits{X.}}
(\byear{2017}).
\btitle{{A Review of Dynamic Network Models with Latent Variables}}.
\bjournal{arXiv e-prints}.
\end{barticle}
\endbibitem

\bibitem[\protect\citeauthoryear{Kivel{\"a} et~al.}{2014}]{Kivela14}
\begin{barticle}[author]
\bauthor{\bsnm{Kivel{\"a}},~\bfnm{Mikko}\binits{M.}},
  \bauthor{\bsnm{Arenas},~\bfnm{Alex}\binits{A.}},
  \bauthor{\bsnm{Barthelemy},~\bfnm{Marc}\binits{M.}},
  \bauthor{\bsnm{Gleeson},~\bfnm{James~P.}\binits{J.~P.}},
  \bauthor{\bsnm{Moreno},~\bfnm{Yamir}\binits{Y.}} \AND
  \bauthor{\bsnm{Porter},~\bfnm{Mason~A.}\binits{M.~A.}}
(\byear{2014}).
\btitle{{Multilayer networks}}.
\bjournal{Journal of Complex Networks}
\bvolume{2}
\bpages{203-271}.
\end{barticle}
\endbibitem

\bibitem[\protect\citeauthoryear{Kumar, Wicker and Swann}{2017}]{Kumar17}
\begin{binproceedings}[author]
\bauthor{\bsnm{Kumar},~\bfnm{Ram Shankar~Siva}\binits{R.~S.~S.}},
  \bauthor{\bsnm{Wicker},~\bfnm{Andrew}\binits{A.}} \AND
  \bauthor{\bsnm{Swann},~\bfnm{Matt}\binits{M.}}
(\byear{2017}).
\btitle{Practical Machine Learning for Cloud Intrusion Detection: Challenges
  and the Way Forward}.
In \bbooktitle{Proceedings of the 10th ACM Workshop on Artificial Intelligence
  and Security}.
\bseries{AISec '17}
\bpages{81--90}.
\end{binproceedings}
\endbibitem

\bibitem[\protect\citeauthoryear{Liben-Nowell and Kleinberg}{2007}]{liben}
\begin{barticle}[author]
\bauthor{\bsnm{Liben-Nowell},~\bfnm{D.}\binits{D.}} \AND
  \bauthor{\bsnm{Kleinberg},~\bfnm{J.}\binits{J.}}
(\byear{2007}).
\btitle{The Link-prediction Problem for Social Networks}.
\bjournal{Journal of the American Society for Information Science and
  Technology}
\bvolume{58}
\bpages{1019--1031}.
\end{barticle}
\endbibitem

\bibitem[\protect\citeauthoryear{{L\"{u}} and Zhou}{2011}]{lu}
\begin{barticle}[author]
\bauthor{\bsnm{{L\"{u}}},~\bfnm{L.}\binits{L.}} \AND
  \bauthor{\bsnm{Zhou},~\bfnm{T.}\binits{T.}}
(\byear{2011}).
\btitle{Link prediction in complex networks: A survey}.
\bjournal{Physica A: Statistical Mechanics and its Applications}
\bvolume{390}
\bpages{1150 - 1170}.
\end{barticle}
\endbibitem

\bibitem[\protect\citeauthoryear{Menon and Elkan}{2011}]{menon}
\begin{bincollection}[author]
\bauthor{\bsnm{Menon},~\bfnm{A.~K.}\binits{A.~K.}} \AND
  \bauthor{\bsnm{Elkan},~\bfnm{C.}\binits{C.}}
(\byear{2011}).
\btitle{Link Prediction via Matrix Factorization}.
In \bbooktitle{Machine Learning and Knowledge Discovery in Databases: European
  Conference, ECML PKDD 2011, Part II}
\bpages{437--452}.
\bpublisher{Springer Berlin Heidelberg}, \baddress{Berlin, Heidelberg}.
\end{bincollection}
\endbibitem

\bibitem[\protect\citeauthoryear{Metelli and Heard}{2019}]{Metelli19}
\begin{barticle}[author]
\bauthor{\bsnm{Metelli},~\bfnm{Silvia}\binits{S.}} \AND
  \bauthor{\bsnm{Heard},~\bfnm{Nicholas~A.}\binits{N.~A.}}
(\byear{2019}).
\btitle{On {Bayesian} new edge prediction and anomaly detection in computer
  networks}.
\bjournal{Annals of Applied Statistics}
\bvolume{13}
\bpages{2586-2610}.
\end{barticle}
\endbibitem

\bibitem[\protect\citeauthoryear{Nakajima, Sugiyama and
  Tomioka}{2010}]{Nakajima10}
\begin{binproceedings}[author]
\bauthor{\bsnm{Nakajima},~\bfnm{S.}\binits{S.}},
  \bauthor{\bsnm{Sugiyama},~\bfnm{M.}\binits{M.}} \AND
  \bauthor{\bsnm{Tomioka},~\bfnm{R.}\binits{R.}}
(\byear{2010}).
\btitle{Global Analytic Solution for Variational Bayesian Matrix
  Factorization}.
In \bbooktitle{Advances in Neural Information Processing Systems 23}
\bpages{1768--1776}.
\end{binproceedings}
\endbibitem

\bibitem[\protect\citeauthoryear{Neil et~al.}{2013}]{neil}
\begin{barticle}[author]
\bauthor{\bsnm{Neil},~\bfnm{J.}\binits{J.}},
  \bauthor{\bsnm{Hash},~\bfnm{C.}\binits{C.}},
  \bauthor{\bsnm{Brugh},~\bfnm{A.}\binits{A.}},
  \bauthor{\bsnm{Fisk},~\bfnm{M.}\binits{M.}} \AND
  \bauthor{\bsnm{Storlie},~\bfnm{C.~B.}\binits{C.~B.}}
(\byear{2013}).
\btitle{Scan Statistics for the Online Detection of Locally Anomalous
  Subgraphs}.
\bjournal{Technometrics}
\bvolume{55}
\bpages{403-414}.
\end{barticle}
\endbibitem

\bibitem[\protect\citeauthoryear{Nguyen and Zhu}{2013}]{Nguyen13}
\begin{barticle}[author]
\bauthor{\bsnm{Nguyen},~\bfnm{Jennifer}\binits{J.}} \AND
  \bauthor{\bsnm{Zhu},~\bfnm{Mu}\binits{M.}}
(\byear{2013}).
\btitle{Content-boosted matrix factorization techniques for recommender
  systems}.
\bjournal{Statistical Analysis and Data Mining: The ASA Data Science Journal}
\bvolume{6}
\bpages{286--301}.
\end{barticle}
\endbibitem

\bibitem[\protect\citeauthoryear{{Papastamoulis} and
  {Ntzoufras}}{2020}]{Papastamoulis20}
\begin{barticle}[author]
\bauthor{\bsnm{{Papastamoulis}},~\bfnm{Panagiotis}\binits{P.}} \AND
  \bauthor{\bsnm{{Ntzoufras}},~\bfnm{Ioannis}\binits{I.}}
(\byear{2020}).
\btitle{{On the identifiability of Bayesian factor analytic models}}.
\bjournal{arXiv e-prints}
\bpages{arXiv:2004.05105}.
\end{barticle}
\endbibitem

\bibitem[\protect\citeauthoryear{Paquet and Koenigstein}{2013}]{Paquet13}
\begin{binproceedings}[author]
\bauthor{\bsnm{Paquet},~\bfnm{U.}\binits{U.}} \AND
  \bauthor{\bsnm{Koenigstein},~\bfnm{N.}\binits{N.}}
(\byear{2013}).
\btitle{One-class Collaborative Filtering with Random Graphs}.
In \bbooktitle{Proceedings of the 22nd International Conference on World Wide
  Web}.
\bseries{WWW '13}
\bpages{999--1008}.
\bpublisher{ACM}, \baddress{New York, NY, USA}.
\end{binproceedings}
\endbibitem

\bibitem[\protect\citeauthoryear{Salakhutdinov and Mnih}{2007}]{prob_mat_fact}
\begin{binproceedings}[author]
\bauthor{\bsnm{Salakhutdinov},~\bfnm{R.}\binits{R.}} \AND
  \bauthor{\bsnm{Mnih},~\bfnm{A.}\binits{A.}}
(\byear{2007}).
\btitle{Probabilistic Matrix Factorization}.
In \bbooktitle{Proceedings of the 20th International Conference on Neural
  Information Processing Systems}.
\bseries{NIPS'07}
\bpages{1257--1264}.
\end{binproceedings}
\endbibitem

\bibitem[\protect\citeauthoryear{Salter-Townshend and Murphy}{2013}]{Salter13}
\begin{barticle}[author]
\bauthor{\bsnm{Salter-Townshend},~\bfnm{Michael}\binits{M.}} \AND
  \bauthor{\bsnm{Murphy},~\bfnm{Thomas~Brendan}\binits{T.~B.}}
(\byear{2013}).
\btitle{Variational Bayesian inference for the Latent Position Cluster Model
  for network data}.
\bjournal{Computational Statistics \& Data Analysis}
\bvolume{57}
\bpages{661--671}.
\end{barticle}
\endbibitem

\bibitem[\protect\citeauthoryear{Schein et~al.}{2015}]{schein}
\begin{binproceedings}[author]
\bauthor{\bsnm{Schein},~\bfnm{A.}\binits{A.}},
  \bauthor{\bsnm{Paisley},~\bfnm{J.}\binits{J.}},
  \bauthor{\bsnm{Blei},~\bfnm{D.~M.}\binits{D.~M.}} \AND
  \bauthor{\bsnm{Wallach},~\bfnm{H.}\binits{H.}}
(\byear{2015}).
\btitle{Bayesian {P}oisson tensor factorization for inferring multilateral
  relations from sparse dyadic event counts}.
In \bbooktitle{Proceedings of the 21th ACM SIGKDD International Conference on
  Knowledge Discovery and Data Mining}
\bpages{1045--1054}.
\bpublisher{ACM}.
\end{binproceedings}
\endbibitem

\bibitem[\protect\citeauthoryear{Schein et~al.}{2016}]{schein2}
\begin{binproceedings}[author]
\bauthor{\bsnm{Schein},~\bfnm{A.}\binits{A.}},
  \bauthor{\bsnm{Zhou},~\bfnm{M.}\binits{M.}},
  \bauthor{\bsnm{Blei},~\bfnm{D.~M.}\binits{D.~M.}} \AND
  \bauthor{\bsnm{Wallach},~\bfnm{H.}\binits{H.}}
(\byear{2016}).
\btitle{Bayesian {P}oisson Tucker decomposition for learning the structure of
  international relations}.
In \bbooktitle{Proceedings of the 33rd International Conference on Machine
  Learning , New York, NY, USA}.
\end{binproceedings}
\endbibitem

\bibitem[\protect\citeauthoryear{Seeger and Bouchard}{2012}]{Seeger12}
\begin{binproceedings}[author]
\bauthor{\bsnm{Seeger},~\bfnm{M.}\binits{M.}} \AND
  \bauthor{\bsnm{Bouchard},~\bfnm{G.}\binits{G.}}
(\byear{2012}).
\btitle{Fast variational {B}ayesian inference for non-conjugate matrix
  factorization models}.
In \bbooktitle{Artificial Intelligence and Statistics}
\bpages{1012--1018}.
\end{binproceedings}
\endbibitem

\bibitem[\protect\citeauthoryear{Sewell and Chen}{2015}]{Sewell15}
\begin{barticle}[author]
\bauthor{\bsnm{Sewell},~\bfnm{Daniel~K.}\binits{D.~K.}} \AND
  \bauthor{\bsnm{Chen},~\bfnm{Yuguo}\binits{Y.}}
(\byear{2015}).
\btitle{Latent Space Models for Dynamic Networks}.
\bjournal{Journal of the American Statistical Association}
\bvolume{110}
\bpages{1646-1657}.
\end{barticle}
\endbibitem

\bibitem[\protect\citeauthoryear{Singh and Gordon}{2008}]{singh}
\begin{binproceedings}[author]
\bauthor{\bsnm{Singh},~\bfnm{A.~P.}\binits{A.~P.}} \AND
  \bauthor{\bsnm{Gordon},~\bfnm{G.~J.}\binits{G.~J.}}
(\byear{2008}).
\btitle{Relational Learning via Collective Matrix Factorization}.
In \bbooktitle{Proceedings of the 14th ACM SIGKDD International Conference on
  Knowledge Discovery and Data Mining}.
\bseries{KDD '08}
\bpages{650--658}.
\bpublisher{ACM}, \baddress{New York, NY, USA}.
\end{binproceedings}
\endbibitem

\bibitem[\protect\citeauthoryear{Turcotte, Kent and Hash}{2018}]{Turcotte18}
\begin{binbook}[author]
\bauthor{\bsnm{Turcotte},~\bfnm{Melissa J.~M.}\binits{M.~J.~M.}},
  \bauthor{\bsnm{Kent},~\bfnm{Alexander~D.}\binits{A.~D.}} \AND
  \bauthor{\bsnm{Hash},~\bfnm{Curtis}\binits{C.}}
(\byear{2018}).
\btitle{Unified Host and Network Data Set}
In \bbooktitle{Data Science for Cyber-Security}
\bchapter{1},
\bpages{1-22}.
\bpublisher{World Scientific}.
\end{binbook}
\endbibitem

\bibitem[\protect\citeauthoryear{Turcotte et~al.}{2016}]{turcotte}
\begin{binproceedings}[author]
\bauthor{\bsnm{Turcotte},~\bfnm{M.}\binits{M.}},
  \bauthor{\bsnm{Moore},~\bfnm{J.}\binits{J.}},
  \bauthor{\bsnm{Heard},~\bfnm{N.~A.}\binits{N.~A.}} \AND
  \bauthor{\bsnm{McPhall},~\bfnm{A.}\binits{A.}}
(\byear{2016}).
\btitle{Poisson factorization for peer-based anomaly detection}.
In \bbooktitle{2016 IEEE Conference on Intelligence and Security Informatics
  (ISI)}
\bpages{208--210}.
\bdoi{10.1109/ISI.2016.7745472}
\end{binproceedings}
\endbibitem

\bibitem[\protect\citeauthoryear{{Wu} et~al.}{2020}]{Wu20}
\begin{barticle}[author]
\bauthor{\bsnm{{Wu}},~\bfnm{Z.}\binits{Z.}},
  \bauthor{\bsnm{{Pan}},~\bfnm{S.}\binits{S.}},
  \bauthor{\bsnm{{Chen}},~\bfnm{F.}\binits{F.}},
  \bauthor{\bsnm{{Long}},~\bfnm{G.}\binits{G.}},
  \bauthor{\bsnm{{Zhang}},~\bfnm{C.}\binits{C.}} \AND
  \bauthor{\bsnm{{Yu}},~\bfnm{P.~S.}\binits{P.~S.}}
(\byear{2020}).
\btitle{A Comprehensive Survey on Graph Neural Networks}.
\bjournal{IEEE Transactions on Neural Networks and Learning Systems}
\bpages{1-21}.
\end{barticle}
\endbibitem

\bibitem[\protect\citeauthoryear{Zhang and Chen}{2018}]{Zhang18}
\begin{bincollection}[author]
\bauthor{\bsnm{Zhang},~\bfnm{Muhan}\binits{M.}} \AND
  \bauthor{\bsnm{Chen},~\bfnm{Yixin}\binits{Y.}}
(\byear{2018}).
\btitle{Link Prediction Based on Graph Neural Networks}.
In \bbooktitle{Advances in Neural Information Processing Systems 31}
\bpages{5165--5175}.
\end{bincollection}
\endbibitem

\bibitem[\protect\citeauthoryear{Zhang and Wang}{2015}]{Zhang15}
\begin{binproceedings}[author]
\bauthor{\bsnm{Zhang},~\bfnm{W.}\binits{W.}} \AND
  \bauthor{\bsnm{Wang},~\bfnm{J.}\binits{J.}}
(\byear{2015}).
\btitle{A Collective {B}ayesian {P}oisson Factorization Model for Cold-start
  Local Event Recommendation}.
In \bbooktitle{Proceedings of the 21th ACM SIGKDD International Conference on
  Knowledge Discovery and Data Mining}
\bpages{1455--1464}.
\end{binproceedings}
\endbibitem

\bibitem[\protect\citeauthoryear{Zhou}{2015}]{zhou}
\begin{binproceedings}[author]
\bauthor{\bsnm{Zhou},~\bfnm{M.}\binits{M.}}
(\byear{2015}).
\btitle{Infinite Edge Partition Models for Overlapping Community Detection and
  Link Prediction}.
In \bbooktitle{Proceedings of the Eighteenth International Conference on
  Artificial Intelligence and Statistics, {AISTATS}}.
\end{binproceedings}
\endbibitem

\end{thebibliography}
\bibliographystyle{imsart-nameyear}

\appendix

\section{Full conditional distributions in the extended PMF model} \label{supp_full}

First note that, conditional on $N_{ij}$, $\mvec Z_{ij}=(Z_{ij1},\dots,Z_{ij(R+KH)})$ has a multinomial distribution, 
$$\mvec Z_{ij}\vert N_{ij},\bm\alpha_i,\bm\beta_j,\bm\Phi \sim\mathrm{Mult}\left(N_{ij},\bm\pi_{ij}\right),$$
where $\bm\pi_{ij}$ is the probability vector proportional to $$(\alpha_{i1}\beta_{j1},\dots,\alpha_{iR}\beta_{jR},\phi_{11}x_{i1}y_{j1},\dots,\phi_{KH}x_{iK}y_{jH}).$$ Therefore, setting $\psi_{ij}=\bm\alpha_i^\top\bm\beta_j+\vec 1_K^\top(\bm\Phi\odot\vec x_i\vec y_j^\top)\vec 1_H$,
\begin{align}
p(N_{ij},\mvec Z_{ij}\vert\bm\alpha_i,\bm\beta_j,\bm\Phi,\mvec A) =\left\{\begin{array}{ll} \mathrm{Pois}_+(\psi_{ij})\mathrm{Mult}(N_{ij},\bm\pi_{ij}) & A_{ij}>0, \\ \delta_0(N_{ij})\delta_{\vec 0}(\mvec Z_{ij}) & A_{ij}=0, \end{array}\right. \label{block_gibbs}
\end{align} 
where $\mathrm{Pois}_+(\cdot)$ denotes the zero-truncated Poisson distribution.
The user and host latent features complete conditionals are gamma, where
\begin{gather}
\alpha_{ir}\vert\bm\beta,\zeta_i^{(\alpha)},\mvec Z \sim\Gamma\left(a^{(\alpha)}+\nlsum_{j=1}^{\abs{V}} Z_{ijr},\zeta_i^{(\alpha)}+\nlsum_{j=1}^{\abs{V}}\beta_{jr}\right), \\ 
\beta_{jr}\vert\bm\alpha,\zeta_j^{(\beta)},\mvec Z \sim\Gamma\left(a^{(\beta)}+\nlsum_{i=1}^{\abs{U}} Z_{ijr}, \zeta_j^{(\beta)} + \nlsum_{i=1}^{\abs{U}} \alpha_{ir}\right),
\end{gather}
and
\begin{gather}
\zeta_i^{(\alpha)}\vert\bm\alpha_i\sim\Gamma\left(b^{(\alpha)}+Ra^{(\alpha)},c^{(\alpha)}+\nlsum_{r=1}^R \alpha_{ir} \right),\\ \zeta_j^{(\beta)}\vert\bm\beta_j\sim\Gamma\left(b^{(\beta)}+Ra^{(\beta)},c^{(\beta)}+\nlsum_{r=1}^R \beta_{jr} \right). \label{zeta_eta}
\end{gather}
Similarly,
\begin{gather}
\phi_{kh}\vert\zeta^{(\phi)},\mvec Z\sim \Gamma\left(a^{(\phi)}+\nlsum_{i=1}^{\abs{U}}\nlsum_{j=1}^{\abs{V}} {Z_{ijl}},\
\zeta^{(\phi)}+\nlsum_{i=1}^{\abs{U}} x_{ik}\nlsum_{j=1}^{\abs{V}} y_{jh}\right), \label{phis} \notag\\
\zeta^{(\phi)}\vert\vec\Phi\sim\Gamma\left(b^{(\phi)}+KHa^{(\phi)},\
c^{(\phi)}+\nlsum_{k=1}^K \nlsum_{h=1}^H\phi_{kh}\right),
\end{gather}
where $l$ is the index corresponding to the covariate pair $(k,h)$. 

\section{Variational inference in the extended PMF model} \label{supp_vi}

As all of the factors in the variational approximation given in \eqref{mean_field} take the same distributional form of the complete conditionals,
\begin{align}
q(N_{ij},\ &\mvec Z_{ij}\vert\theta_{ij},\bm\chi_{ij}) =
\left\{\begin{array}{ll} \mathrm{Pois}_+(\theta_{ij})\mathrm{Mult}(N_{ij},\bm\chi_{ij}) & A_{ij}>0, \\ \delta_0(N_{ij})\delta_{\vec 0}(\mvec Z_{ij}) & A_{ij}=0. \end{array}\right.  \label{factorisation}
\end{align}
Let $\psi_{ij}$ denote the rate $\sum_{r=1}^R\alpha_{ir}\beta_{jr}+\sum_{k=1}^K\sum_{h=1}^H\phi_{kh}x_{ik}y_{jh}$ of the Poisson distribution for $N_{ij}$, and
$\psi_{ijl},\ l=1,\dots,R+KH$, represent the individual elements in the sum. 
To get the update equations for $\bm\theta$ and $\bm\chi$, following \eqref{cavi_update}, for $A_{ij}>0$,
\begin{equation}
\mathbb E^q_{-N_{ij},\mvec Z_{ij}}\left\{ \log p(N_{ij},\mvec Z_{ij}\vert \bm\alpha_i,\bm\beta_j,\bm\Phi) \right\} = \sum_l \left\{ Z_{ijl}\mathbb E^q_{-N_{ij},\mvec Z_{ij}}\left(\log\psi_{ijl}\right) -\log(Z_{ijl}!)\right\} + k, \label{expect_nij}
\end{equation}
where $k$ is a constant with respect to $N_{ij}$ and $\mvec Z_{ij}$. Hence:
\begin{equation}
q^\star(N_{ij},\mvec Z_{ij}) \propto \nlprod_{l=1}^{R+KH} \exp\left\{ \mathbb E^q_{-N_{ij},\mvec Z_{ij}}(\log\psi_{ijl}) \right\}^{Z_{ijl}} \Big/ Z_{ijl}!,
\end{equation}
with domain of $\mvec Z_{ij}$ constrained to have $\sum_l Z_{ijl}>0$. Multiplying and dividing the expression by $N_{ij}!$ and $[\sum_l \exp\{ \mathbb E^q_{-N_{ij},\mvec Z_{ij}}(\log\psi_{ijl})\}]^{N_{ijl}}$ gives
a distribution
which has the same form of \eqref{factorisation}. 
Therefore the rate $\theta_{ij}$ of the zero truncated Poisson is updated using $\sum_l \exp\{ \mathbb E^q_{-N_{ij},\mvec Z_{ij}}(\log\psi_{ijl})\}$, see step 5 in Algorithm~\ref{algo_vi} for the resulting final expression. The update for the vector of probabilities $\bm\chi_{ij}$ is given by a slight extension of the standard result for variational inference in the PMF model \citep{gopalan} to include the covariate terms, see step 6 of Algorithm~\ref{algo_vi}. The remaining updates are essentially analogous to standard PMF \citep{gopalan}. 

\section{Inference in the seasonal model} \label{supp_seasonal}

The inferential procedure for the seasonal model follows the same guidelines used for the non-seasonal model. Given the unobserved count $N_{ijt}$, latent variables $Z_{ijtl}$ are added, representing the contribution of the component $l$ to the total count $N_{ijt}$: $N_{ijt}=\sum_{l} Z_{ijtl}$. The full conditional for $N_{ijt}$ and $Z_{ijt}$ follows \eqref{block_gibbs}, except the rate for the Poisson and probability vectors for the multinomial will now depend on the seasonal parameters $\gamma_{itr}$, $\delta_{jtr}$, and $\omega_{kht}$. Letting $p$ denote a seasonal segment in $\{1,\ldots,P\}$ the full conditionals for the rate parameters are:
\begin{gather}
\alpha_{ir}\vert\mvec Z,\bm\beta,\bm\gamma,\bm\delta,\zeta_i^{(\alpha)} \overset{d}{\sim}\Gamma\left(a^{(\alpha)}+\sum_{j=1}^{\abs{V}}\sum_{t=1}^T Z_{ijtr}, \zeta_i^{(\alpha)}+\sum_{t=1}^T\gamma_{itr}\sum_{j=1}^{\abs{V}}\beta_{jr}\delta_{jtr} \right), \\
\gamma_{ipr}\vert\mvec Z,\bm\alpha,\bm\beta,\bm\delta,\zeta_p^{(\gamma)} \overset{d}{\sim}\Gamma\left(a^{(\gamma)}+\sum_{j=1}^{\abs{V}}\sum_{t:t=p} Z_{ijtr}, \zeta_p^{(\gamma)}+\alpha_{ir}\sum_{j=1}^{\abs{V}} \beta_{jr}\sum_{t:t=p} \delta_{jt r} \right), \notag \\ 
\phi_{kh}\vert\mvec Z,
\zeta^{(\phi)} \overset{d}{\sim}\Gamma\left(a^{(\phi)}+\sum_{i=1}^{\abs{U}}\sum_{j=1}^{\abs{V}}\sum_{t=1}^T Z_{ijtl}, \zeta^{(\phi)} +T\tilde x_k\tilde y_h
\right), 
\end{gather}
where $\tilde x_k=\sum_{i=1}^{\abs{U}} x_{ik}$ and $\tilde y_h=\sum_{j=1}^{\abs{V}} y_{jh}$. Similar results are available for $\beta_{jr}$ and $\delta_{jpr}$. 
Also:
\begin{gather}
\zeta_p^{(\gamma)}\vert\bm\gamma\overset{d}{\sim}\Gamma\left(b^{(\gamma)}+\abs{U}Ra^{(\gamma)},c^{(\gamma)}+\sum_{i=1}^{\abs{U}}\sum_{r=1}^R \gamma_{ipr} \right),
\end{gather}
and similarly for $\zeta_p^{(\delta)}$. For $\zeta_i^{(\alpha)}$ and $\zeta_j^{(\beta)}$, the conditional distribution is equivalent to \eqref{zeta_eta}. The mean-field variational family is again used implying a factorisation similar to \eqref{mean_field}, so that
\begin{align}
  q(&\bm\alpha,\bm\beta,\bm\Phi,\bm\gamma,\bm\delta,
  \bm\zeta,\mvec N,\mvec Z) = \nlprod_{i,j,t}q(N_{ijt},\mvec Z_{ijt}\vert\theta_{ijt},\bm\chi_{ijt}) \times \nlprod_{i,r} q(\alpha_{ir}\vert\lambda_{ir}^{(\alpha)},\mu_{ir}^{(\alpha)}) \\ 
  & \times\nlprod_{j,r}q(\beta_{jr}\vert\lambda_{jr}^{(\beta)},\mu_{jr}^{(\beta)})\times\nlprod_{k,h}q(\phi_{kh}\vert\lambda_{kh}^{(\phi)},\mu_{kh}^{(\phi)}) \times\nlprod_iq(\zeta_i^{(\alpha)}\vert\nu_i^{(\alpha)},\xi_i^{(\alpha)}) \\
  & \times\nlprod_jq(\zeta_j^{(\beta)}\vert\nu_j^{(\beta)},\xi_j^{(\beta)}) \times q(\zeta^{(\phi)}\vert\nu^{(\phi)},\xi^{(\phi)}) \times\nlprod_{i,q,r}q(\gamma_{ipr}\vert\lambda_{ipr}^{(\gamma)},\mu_{ipr}^{(\gamma)}) \\ 
  &\times\nlprod_{j,p,r}q(\delta_{jpr}\vert\lambda_{jpr}^{(\delta)},\mu_{jpr}^{(\delta)}) 
\times \nlprod_p q(\zeta_p^{(\gamma)}\vert\nu_p^{(\gamma)},\xi_p^{(\gamma)})q(\zeta_p^{(\delta)}\vert\nu_p^{(\delta)},\xi_p^{(\delta)}).
\end{align}

As in Section~\ref{inference}, 
each $q(\cdot)$ has the same form of the full conditional distributions for the corresponding parameter or group of parameters. Again the variational parameters are updated using CAVI and a similar algorithm is obtained to that detailed in Algorithm~\ref{algo_vi}, where steps 7, 8, 9 and 10 are modified to include the time dependent parameters. It follows that for the user-specific parameters the update equations take the form:
\begin{gather}
\lambda_{ir}^{(\alpha)} = a^{(\alpha)}+\sum_{j=1}^{\abs{V}}\sum_{t=1}^T \frac{A_{ijt}\theta_{ijt}\chi_{ijtr}}{1-e^{-\theta_{ijt}}},\  
\mu_{ir}^{(\alpha)} = \frac{\nu_i^{(\alpha)}}{\xi_i^{(\alpha)}}+\sum_{t=1}^T\frac{\lambda_{it r}^{(\gamma)}}{\mu_{it r}^{(\gamma)}}\sum_{j=1}^{\abs{V}}\frac{\lambda_{jr}^{(\beta)}}{\mu_{jr}^{(\beta)}}\frac{\lambda_{jt r}^{(\delta)}}{\mu_{jt r}^{(\delta)}}, \notag \\
\lambda_{ipr}^{(\gamma)}=a^{(\gamma)}+\sum_{j=1}^{\abs{V}}\sum_{t:t=p} \frac{A_{ijt}\theta_{ijt}\chi_{ijtr}}{1-e^{-\theta_{ijt}}},\ 
\mu_{ipr}^{(\gamma)} =  \frac{\nu_p^{(\gamma)}}{\xi_p^{(\gamma)}}+\frac{\lambda_{ir}^{(\alpha)}}{\mu_{ir}^{(\alpha)}}\sum_{j=1}^{\abs{V}} \frac{\lambda_{jr}^{(\beta)}}{\mu_{jr}^{(\beta)}}\sum_{t:t =p} \frac{\lambda_{jt r}^{(\delta)}}{\mu_{jt r}^{(\delta)}},
\end{gather}
and similar results can be obtained for the host-specific parameters $\lambda_{ir}^{(\beta)}$, $\mu_{jr}^{(\beta)}$, $\lambda_{jpr}^{(\delta)}$ and $\mu_{jpr}^{(\delta)}$. The updates for $\nu_i^{(\alpha)},\  \xi_i^{(\alpha)},\ \nu_j^{(\beta)}$ and $\xi_j^{(\beta)}$ are identical to steps 7 and 8 in Algorithm~\ref{algo_vi}. For the covariates
\begin{gather}
\lambda_{kh}^{(\phi)}=a^{(\phi)}+\sum_{i=1}^{\abs{U}}\sum_{j=1}^{\abs{V}}\sum_{t=1}^T \frac{A_{ijt}\theta_{ijt}\chi_{ijtl}}{1-e^{-\theta_{ijt}}},\
\mu_{kh}^{(\phi)} =  \frac{\nu^{(\phi)}}{\xi^{(\phi)}}+\tilde x_k\tilde y_hT. 
\end{gather} 
The updates for $\nu^{(\phi)}$ and $\xi^{(\phi)}$ are the same as step 9 in Algorithm~\ref{algo_vi}. Finally, for the time dependent hyperparameters: 
\begin{align}
\nu_p^{(\gamma)} = b^{(\gamma)} + \abs{U}Ra^{(\gamma)},\ 
\xi_p^{(\gamma)} = c^{(\gamma)} + \sum_{i=1}^{\abs{U}}\sum_{r=1}^R \frac{\lambda_{ipr}^{(\gamma)}}{\mu_{ipr}^{(\gamma)}},
\end{align}
and similarly for $\nu_p^{(\delta)}$ and $\xi_p^{(\delta)}$.

The updates for $\theta_{ijt}$ and $\bm\chi_{ijt}$ are similar to Appendix~\ref{supp_vi}. 
An expansion similar to \eqref{expect_nij} can be applied to the expectation $\mathbb E^q_{-N_{ijt},\mvec Z_{ijt}}\{\log p(N_{ijt},\mvec Z_{ijt}\vert \bm\alpha_i,\bm\beta_j,\bm\gamma_{it },\bm\delta_{jt },\bm\Phi
\}$, and the update equations for $\theta_{ijt}$ and $\chi_{ijtr}$ can be derived similarly to Appendix~\ref{supp_vi}:
\begin{align}
  \theta_{ijt} =&\ \sum_{r=1}^R \exp\left\{\Psi(\lambda_{ir}^{(\alpha)})-\log(\mu_{ir}^{(\alpha)})+\Psi(\lambda_{jr}^{(\beta)})-\log(\mu_{jr}^{(\beta)})\right. \\
    &\hspace{2.3cm} \left. +\ \Psi(\lambda_{it r}^{(\gamma)})-\log(\mu_{it r}^{(\gamma)})+\Psi(\lambda_{jt r}^{(\delta)})-\log(\mu_{jt r}^{(\delta)})\right\} \\
    &+\sum_{k=1}^K\sum_{h=1}^H x_{ik}y_{jh}\exp\left\{\Psi(\lambda_{kh}^{(\phi)})-\log(\mu_{kh}^{(\phi)})
    \right\}, \\[1em]
  \chi_{ijtl}\propto&\left\{ \begin{array}{ll} \exp\left\{\Psi(\lambda_{il}^{(\alpha)})-\log(\mu_{il}^{(\alpha)})+\Psi(\lambda_{jl}^{(\beta)})-\log(\mu_{jl}^{(\beta)})\right. &  \\
    \hspace{.8cm}\left.\ +\ \Psi(\lambda_{it l}^{(\gamma)})-\log(\mu_{it l}^{(\gamma)})+\Psi(\lambda_{jt l}^{(\delta)})-\log(\mu_{jt l}^{(\delta)})\right\} &l\leq R, \\
  x_{ik}y_{jh}\exp\left\{\Psi(\lambda_{kh}^{(\phi)})-\log(\mu_{kh}^{(\phi)})
  \right\} & l>R.
  \end{array}\right. 
\end{align}

\section{Inference in the joint model} \label{supp_joint}

Variational inference for the joint PMF model presented in Section~\ref{joint_model} proceeds similarly to Algorithm~\ref{algo_vi}. 
The conditional posterior distributions are essentially the same as PMF and EPMF. 
The only exception is the conditional posterior distribution of $\alpha_{ir}$:
\begin{align}
\alpha_{ir}\vert\mvec Z,\mvec Z^\prime, \bm\beta, \bm\beta^\prime, \zeta_i^{(\alpha)} \sim \Gamma\left(a^{(\alpha)} + \sum_{j=1}^{\abs{V}} Z_{ijr} + \sum_{j=1}^{\abs{V^\prime}} Z_{ijr}^\prime,\ \zeta_i^{(\alpha)} + \sum_{j=1}^{\abs{V}} \beta_{jr} + \sum_{j=1}^{\abs{V^\prime}} \beta_{jr}^\prime \right).
\end{align}
For variational inference, a factorisation similar to \eqref{mean_field} is assumed, further multiplied by the approximation for the posteriors of the additional components of the model: mainly $q(\beta_{jr}^\prime\vert\lambda_{jr}^{(\beta^\prime)},\mu_{jr}^{(\beta^\prime)})$, $q(\phi_{kh}^\prime\vert\lambda_{kh}^{(\phi^\prime)},\mu_{kh}^{(\phi^\prime)})$, $q(N_{ij}^\prime,\mvec Z_{ij}^\prime\vert\theta_{ij}^\prime,\bm\chi_{ij}^\prime)$.
Therefore variational inference is also essentially unchanged, except the CAVI update for the parameters of variational approximation for the components $\alpha_{ir}$, which become:
  \begin{align}
 \lambda_{ir}^{(\alpha)} = a^{(\alpha)} +\sum_{j=1}^{\abs{V}} \frac{A_{ij}\theta_{ij}\chi_{ijr}}{1-e^{-\theta_{ij}}} + \sum_{j^\prime=1}^{\abs{V^\prime}} \frac{A_{ij}^\prime\theta_{ij}^\prime\chi_{ijr}^\prime}{1-e^{-\theta_{ij}^\prime}},\ & & 
\mu_{ir}^{(\alpha)} =  \frac{\nu_i^{(\alpha)}}{\xi_i^{(\alpha)}} + \sum_{j=1}^{\abs{V}} \frac{\lambda_{jr}^{(\beta)}}{\mu_{jr}^{(\beta)}} + \sum_{j^\prime=1}^{\abs{V^\prime}} \frac{\lambda_{jr}^{(\beta^\prime)}}{\mu_{jr}^{(\beta^\prime)}}.
  \end{align} 
 Updates for the approximations of the additional components $N_{ij}^\prime, Z_{ijr}^\prime, \beta_{jr}^\prime$ and $\phi_{kh}^\prime$ follow from the updates for the variational parameters $\theta_{ij}, \chi_{ijr}, \lambda_{jr}^{(\beta)}, \mu_{jr}^{(\beta)}, \lambda_{kh}^{(\phi)}$ and $ \mu_{kh}^{(\phi)}$ in Algorithm~\ref{algo_vi}.

\vspace{.5cm}
\footnotesize
\begin{minipage}[t]{0.57\textwidth}
\textsc{Francesco Sanna Passino, Nicholas A. Heard\\[.25em] 
Department of Mathematics\\
Imperial College London\\
180 Queen's Gate\\ 
SW7 2AZ London (United Kingdom)} \\[.25em] 
\textit{E-mail:} {\color{blue}\texttt{f.sannapassino@imperial.ac.uk}} \\
\hspace*{1cm} {\color{blue}\texttt{n.heard@imperial.ac.uk}}
\end{minipage}
\begin{minipage}[t]{0.42\textwidth}
\textsc{Melissa J.\hspace{0.1em}M. Turcotte \\[.25em] 
Microsoft 365 Defender \\
Microsoft Corporation \\
One Microsoft Way \\ 
Redmond, WA 98052 (USA)} \\[.25em] 
\textit{E-mail:} {\color{blue}\texttt{melissa.turcotte@microsoft.com}}
\end{minipage}

\end{document}